\documentclass[prb,superscriptaddress,twocolumn]{revtex4-2}

\usepackage{graphicx,dcolumn,bm,color,amsmath,amssymb,physics}
\usepackage{empheq,dsfont,diagbox,ulem}
\usepackage[theorems,skins]{tcolorbox}
\usepackage{slashed}
\usepackage{tikz} 

\newcommand{\nn}{\nonumber \\}

\begin{document}

\title{Multi-orbital effects on superconductivity in kagome metals: Parquet renormalization group analysis}
\author{Jae-Ho Han}
\email{jaehohan@kaist.ac.kr}
\affiliation{Department of Physics, Korea Advanced Institute of Science and Technology (KAIST), Daejeon 34141, Korea}

\author{SungBin Lee}
\email{sungbin@kaist.ac.kr}
\affiliation{Department of Physics, Korea Advanced Institute of Science and Technology (KAIST), Daejeon 34141, Korea}

\begin{abstract}
    The Van Hove singularities (VHSs), where the electronic density of states diverges due to saddle points in the band structure, play a crucial role in enhancing electronic correlations and driving various instabilities. In particular, VHS-induced superconductivity has earned significant attention due to its potential to achieve high transition temperatures and its tendency to favor exotic pairing states beyond conventional electron-phonon mechanisms. Despite extensive research on VHS-driven superconductivity, the multi-orbital effect on such systems remains less explored. Motivated by recent experiments on several kagome metals under doping and pressure [Z.Zhang {\it et al.}, Phys.Rev.B {\bf 103}, 224513 (2021), Y.Sur {\it et al.}, Nat.Commun. {\bf 14}, 3899 (2023)], we explore the effects of multi-orbital physics and strong correlations induced by VHS in the kagome lattice, focusing on their impact on superconductivity. Using parquet renormalization group analysis, we uncover eight distinct superconducting instabilities, characterized by order parameters with mixed orbital degrees of freedom. Among these, we identify a parameter regime where $d$-wave-like orbital-singlet spin-triplet order parameters dominate as the leading instability. The degenerate spin-triplet states in this regime are capable of breaking time-reversal symmetry, which is a multi-orbital analogue of chiral spin-triplet superconductivity. These findings highlight the interplay between multi-orbital effects on superconductivity and can apply to the kagome metal systems such as $A$V$_3$Sb$_5$ ($A$ = K, Rb, Cs) family.
\end{abstract}

\maketitle

\section{Introduction}

Strong correlations in multi-orbital systems are particularly intriguing, as they give rise to a variety of emergent exotic phases and have been the focus of extensive research. One of the prominent mechanisms for achieving strong electronic correlations is the van Hove singularity (VHS), a point in momentum space where the density of states diverges~\cite{Ashcroft76}. VHSs naturally emerge in systems with triangular lattice structures, such as kagome lattices, graphene, and twisted bilayer graphene. These systems provide an ideal platform for exploring the interplay between strong correlations and lattice geometry, particularly in the contexts of charge-density-wave (CDW), nematicity, and superconductivity.

A notable realization of the kagome lattice in material aspects is found in the family of compounds $A$V$_3$Sb$_5$ ($A$ = K, Rb, Cs)~\cite{Jiang2021,Li2022,Liu2021,Frassineti2023,Song2023}. In these materials, vanadium atoms form a kagome network, and multiple VHSs are situated near the Fermi level, significantly influencing their electronic properties. Recent discoveries have shown that AV$_3$Sb$_5$ systems exhibit CDW orders of 2$\times$2 star-of-David or tri-hexagonal type, driven by VHS-enhanced interactions~\cite{Jiang2021,Li2022,Liu2021,Frassineti2023,Song2023}. These orders, described as triple-$\bm{q}$ bond orders with even parity modulation, highlight the intricate interplay between electronic, orbital, and lattice degrees of freedom. Beyond the conventional CDW, time-reversal-symmetry-breaking (TRSB) phases emerges, leading the anomalous Hall effect without net magnetization~\cite{Yang2024,Wang2023a,Guo2022}. In addition, nematic electronic states can also be developed~\cite{Li2022,Xu2022,Nie2022,Tazai2023,Asaba2024}.

In the context of superconductivity, VHS-induced pairing has been widely explored, uncovering a diverse range of exotic superconducting states beyond conventional electron-phonon mechanisms. These include spin-triplet superconducting phases, as well as chiral phases which break time-reversal symmetry, arising due to enhanced electronic correlations at VHS points. 
In the $A$V$_3$Sb$_5$ family, superconducting phases have been observed in all compounds~\cite{Ortiz2020,Yin2021,Ortiz2021}. Experimental studies have reported unconventional features such as pair-density waves~\cite{Chen2021}, nematicity~\cite{Xiang2021,Ni2021}, and TRSB superconductivity~\cite{Khasanov2022,Guguchia2023,Graham2024}. 
Theoretically, chiral $d+id$ order parameters have been proposed, which are consistent with the absence of nodal gap structures~\cite{Roppongi2023}. Notably, these predictions are largely based on single-orbital theoretical analyses.

While much effort has been devoted to understanding the role of a single VHS in superconductivity, the effects of multiple VHSs, which naturally emerge in kagome metals and other correlated systems, have been studied in different contexts~\cite{Scammell2023}. However, their direct influence on superconducting instabilities remains less explored.
The presence of multiple VHSs introduces additional complexity, including enhanced competition between different pairing channels and the potential for novel multi-component superconducting states. Furthermore, the interplay between VHSs and multi-orbital interactions remains an open question, as multi-orbital degrees of freedom can significantly modify the effective DOS, hybridization, and pairing symmetries.

In terms of experimental aspects, recent experiments have shown that the nature of these VHSs can be tuned through external parameters such as pressure~\cite{Chen2021a,Ni2021,Zhang2021} and chemical substitution~\cite{Li2022a,Sur2023,Song2024,Yousuf2024}. For instance, applying pressure shifts the VHSs closer to or farther from the Fermi level, leading to significant Fermi surface reconstruction and modifications in electronic interactions. Additionally, substituting other transition metals for the vanadium ion effectively introduces electron or hole doping, altering the chemical potential near multiple VHSs. One particularly intriguing aspect of pressure and doping is the reemergence of the superconducting phase~\cite{Zhang2021,Sur2023}, suggesting complex electronic correlation effects. These findings naturally prompt consideration of how multi-orbital effects at VHSs play a crucial role in this reentrant superconducting behavior, potentially stabilizing novel pairing mechanisms and unconventional superconducting states.

In this work, we focus on the kagome lattice as a model system to explore the effects of multi-orbital interactions and VHS on superconductivity. Utilizing the parquet renormalization group ($p$RG) method~\cite{Furukawa1998a,Furukawa1998,Binz2002,Nandkishore2012,Lin2019,Lin2020,Scammell2023}, we systematically investigate the possible superconducting order parameters and construct tentative phase diagrams as functions of the bare interaction strengths. By extending previous studies, we identify parameter regimes where spin-triplet superconductivity emerges and uncover the intricate interplay of correlation effects in multiband kagome systems. Our findings offer valuable insights into the mechanisms underlying unconventional superconductivity, paving the way for further theoretical exploration and experimental validation in various kagome metallic systems.

The structure of this paper is as follows. We begin by introducing the theoretical model for the kagome lattice, outlining the multi-orbital interactions, and specifying the parameters employed in our $p$RG calculations. Next, we provide a detailed explanation of the $p$RG procedure, describing the methodology for tracking the flow of interaction vertices and identifying the dominant instability channels. We then present the results of our $p$RG analysis, including the phase diagram of potential superconducting states and the identification of phases characterized by odd-parity pairing. Finally, we discuss the broader implications of these findings for understanding superconductivity in kagome metals and propose directions for experimental validation and further theoretical exploration.

\section{Model}

\begin{figure}[t!]
\includegraphics[width=\linewidth]{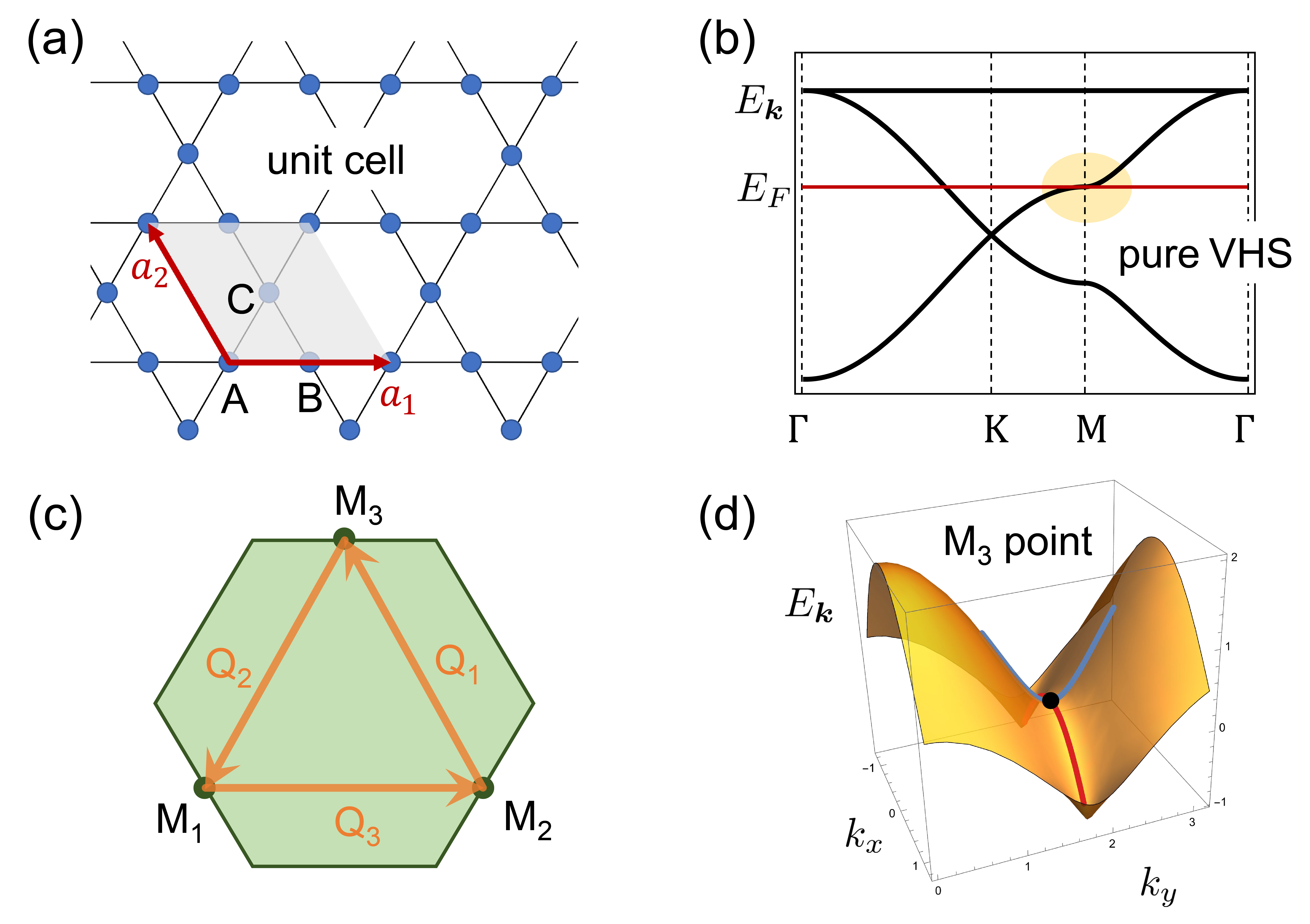}
\caption{(a) The kagome lattice structure. $a_1$ and $a_2$ (red arrows) are hexagonal lattice vectors, and the shaded parallelogram is a unit cell. Three basis points are denoted as $A$, $B$, and $C$. (b) The band structure of the nearest-neighbor tight-binding model is shown. The van Hove singularities (VHSs) are at $M$-points, highlighted in orange oval. The Fermi energy is set near the pure VHSs. (c) Brillouin zone of the hexagonal lattice. The three different $M$-points are connected by vectors, $\bm Q_\alpha$, $\alpha=1,2,3$. (d) Energy dispersion at the VHS is shown, with $M_3$ as an example. The curvatures of the dispersion along the $k_x$ (red line) and $k_y$ (blue line) are opposite, demonstrating its saddle point nature.
}
\label{fig:kagome}
\end{figure}

\begin{figure}[t!]
\includegraphics[width=\linewidth]{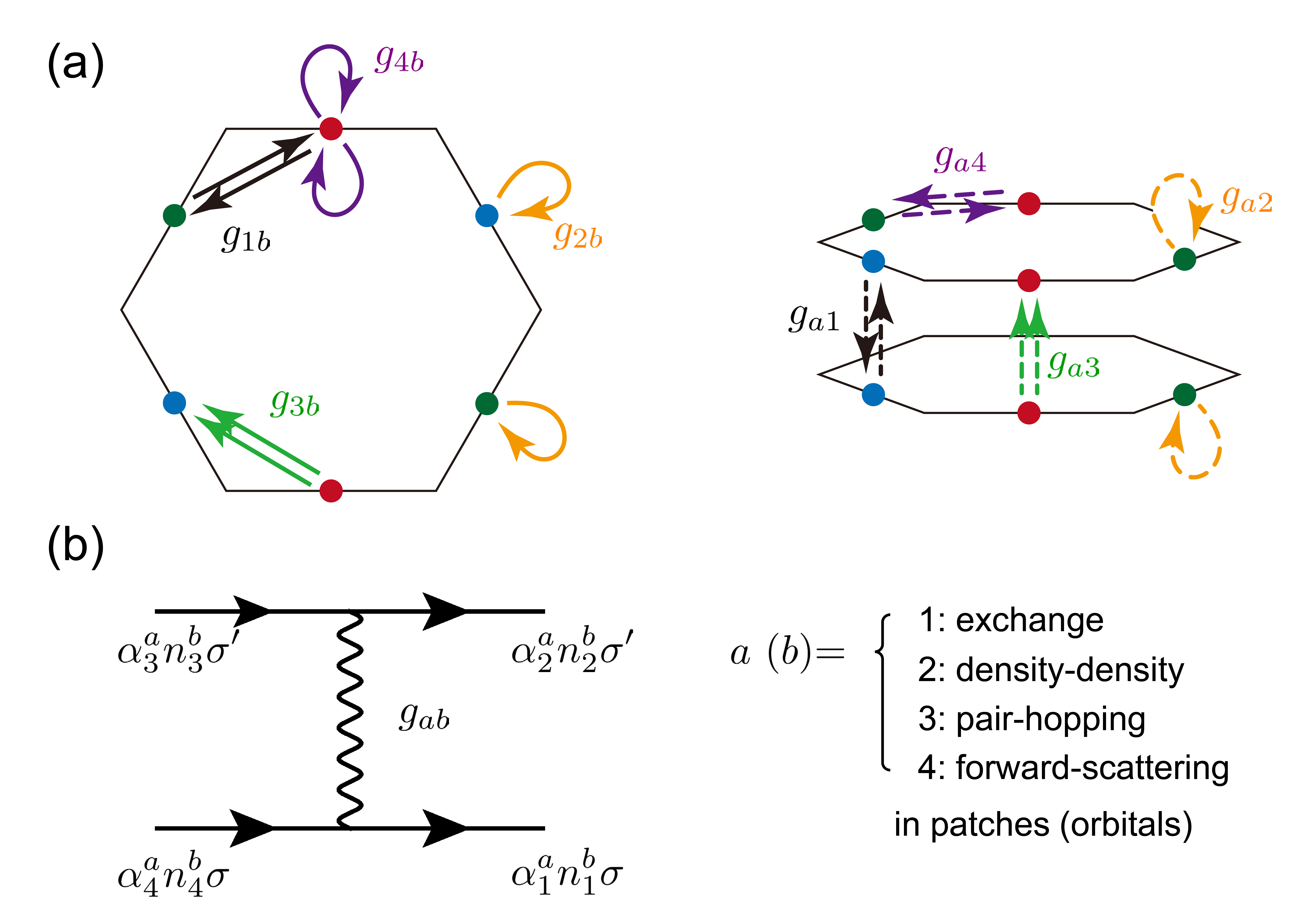}
\caption{(a) The left panel shows intra-orbital interactions: exchange ($g_{1b}$), density-density ($g_{2b}$), pair-hopping ($g_{3b}$), and foward-scattering ($g_{4b}$) interactions. The right panel shows inter-orbital interactions in the same order in orbital space. For concreteness, we show $g_{41}$, $g_{42}$, $g_{43}$, and $g_{14}$ in the right panel. (b) Interaction diagram with corresponding indices.
}
\label{fig:interaction}
\end{figure}

The model consists of electrons with two orbital degrees of freedom on the kagome lattice and their interactions. The free part is a tight-binding Hamiltonian hopping through nearest neighbor sites of kagome lattices as shown in Fig.~\ref{fig:kagome}(a). We introduce orbital degrees of freedom at each lattice site, which is identical for simplicity. Thus, the Hamiltonian for the free part is 
\begin{eqnarray}
    H = -t \sum_{n=1,2} \sum_{\left< ij \right>, \sigma} c_{in\sigma}^\dagger c_{jn\sigma} + h.c.,
\end{eqnarray}
where $c_{in\sigma}$ ($c_{in\sigma}^\dagger$) is creation (annihilation) operator of orbital $n$ and spin $\sigma$ on lattice site $i$, and $\left< ij \right>$ denotes nearest-neighboring sites. The band structure of the Hamiltonian is shown in Fig.~\ref{fig:kagome}(b). The Fermi energy is set to be energy at pure VHSs, which are three $M$ points in the Brillouin zone [Fig.~\ref{fig:kagome}(c), (d)]. To study low-energy physics, we adopt patch description. The free part of the Hamiltonian in patches is
\begin{eqnarray}
    H = \sum_{\alpha = 1, 2, 3} \sum_{n=1,2} \sum_{\sigma} \int_{\bm k} \epsilon_{\alpha}(\bm k) c_{n\alpha\sigma}^\dagger(\bm k) c_{n\alpha\sigma} (\bm k),
\end{eqnarray}
where $\int_{\bm k} = \int\!d^2k/(2\pi)^2$ and $\alpha$ denotes patches as shown in Fig~\ref{fig:interaction}(c). The dispersion of each patch is $\epsilon_1(\bm k) = \frac{t}{2} k_x (k_x + \sqrt 3 k_y)$, $\epsilon_2(\bm k) = -\frac{t}{4} (k_x^2 - 3k_y^2)$, and $\epsilon_3(\bm k) = \frac{t}{2} k_x (k_x - \sqrt 3 k_y)$, respectively [Fig.~\ref{fig:interaction}(d)].

\begin{table}[b!]
\caption{\label{tab:label}
The label of interactions \{$\alpha_i^a$\} = ($\alpha_1^a$, $\alpha_2^a$, $\alpha_3^a$, $\alpha_4^a$) and \{$n_i^b$\} = ($n_1^b$, $n_2^b$, $n_3^b$, $n_4^b$). The processes are exchange (ex), density-density (dd), pair-hopping (ph), and forward scattering (fs). Here, $\alpha$, $\beta$ = 1, 2, 3 with $\alpha \neq \beta$, and $n$=1, 2, $\bar n$ = 2, 1, correspondingly.}
\begin{ruledtabular}
    \begin{tabular}{l |cccc}
     $a$ or $b$ & 1 (ex) & 2 (dd) & 3 (ph) & 4 (fs) \\
    \hline
    \{$\alpha_i^a$\} & ($\alpha$, $\beta$, $\alpha$, $\beta$)  & ($\alpha$, $\beta$, $\beta$, $\alpha$) & ($\alpha$, $\alpha$, $\beta$, $\beta$) & ($\alpha$, $\alpha$, $\alpha$, $\alpha$) \\
    \{$n_i^b$\} & ($n$, $\bar n$, $n$, $\bar n$) & ($n$, $\bar n$, $\bar n$, $n$) & ($n$, $n$, $\bar n$, $\bar n$) & ($n$, $n$, $n$, $n$) \\
    \end{tabular}
\end{ruledtabular}
\end{table}

For the interaction, we consider on-site and nearest-neighboring Coulomb interactions $U$ and $V$. Projecting onto the states within the patches~\cite{Wu2021,Scammell2023}, the interaction can be described by 16 kinds of couplings as
\begin{eqnarray}
    &&H_{int} = \frac{1}{2} \sum_{a,b=1}^4 \sum_{\{\alpha_i^a\}} \sum_{\{n_i^b\}} \int_{\bm k_1} \int_{\bm k_2} \int_{\bm k_3} \nn
    &&g_{ab} c_{\alpha_1^a n_1^b \sigma}^\dagger(\bm k_1) c_{\alpha_2^a n_2^b \sigma'}^\dagger(\bm k_2) c_{\alpha_3^a n_3^b \sigma'}(\bm k_3) c_{\alpha_4^a n_4^b \sigma}(\bm k_4).
\end{eqnarray}
Here, $\sum_{i=1}^4 \bm k_i=0$, $a = 1,2,3,4$ ($b = 1,2,3,4$) denote exchange, density-density, pair-hopping, and forward-scattering process in patches (in orbitals), respectively. $\alpha_i^a$, $n_i^b$ are patches and orbitals of a corresponding process $a$ and $b$. For example, $(a,b) = (1,2)$ is exchange in patches and density-density interaction in orbitals and $\{\alpha_i^1\} = (\alpha,\beta, \alpha, \beta)$ with $\alpha\neq \beta$ (Here, $\alpha$, $\beta$ = 1,2,3 for patches) and $\{n_i^2\} = (n, \bar n, \bar n, n)$ where $\bar n = 2,1$ for $n=1,2$ (Here, $n$, $\bar n$ = 1,2 for orbitals). The labels are summarized in Table~\ref{tab:label}. The nesting vectors that connect the VHSs are $\bm Q_\alpha = \epsilon_{\alpha\beta\gamma} (\bm M_\beta - \bm M_\gamma)$. 

\section{Parquet Renormalization Group Analysis}

\subsection{General formulation}

Here, we perform the standard $p$RG analysis~\cite{Furukawa1998a,Furukawa1998,Binz2002,Nandkishore2012,Lin2019,Lin2020,Scammell2023}. In the conventional Wilsonian RG approach, the energy modes are separated into high- and low-energy components, and the high-energy modes are integrated out. This process generates interactions among the low-energy modes and renormalizes the coupling constants. By rescaling the energy cutoff, the resulting action can be compared with the original one, allowing the derivation of the beta functions.

\begin{figure}[t!]
\includegraphics[width=\linewidth]{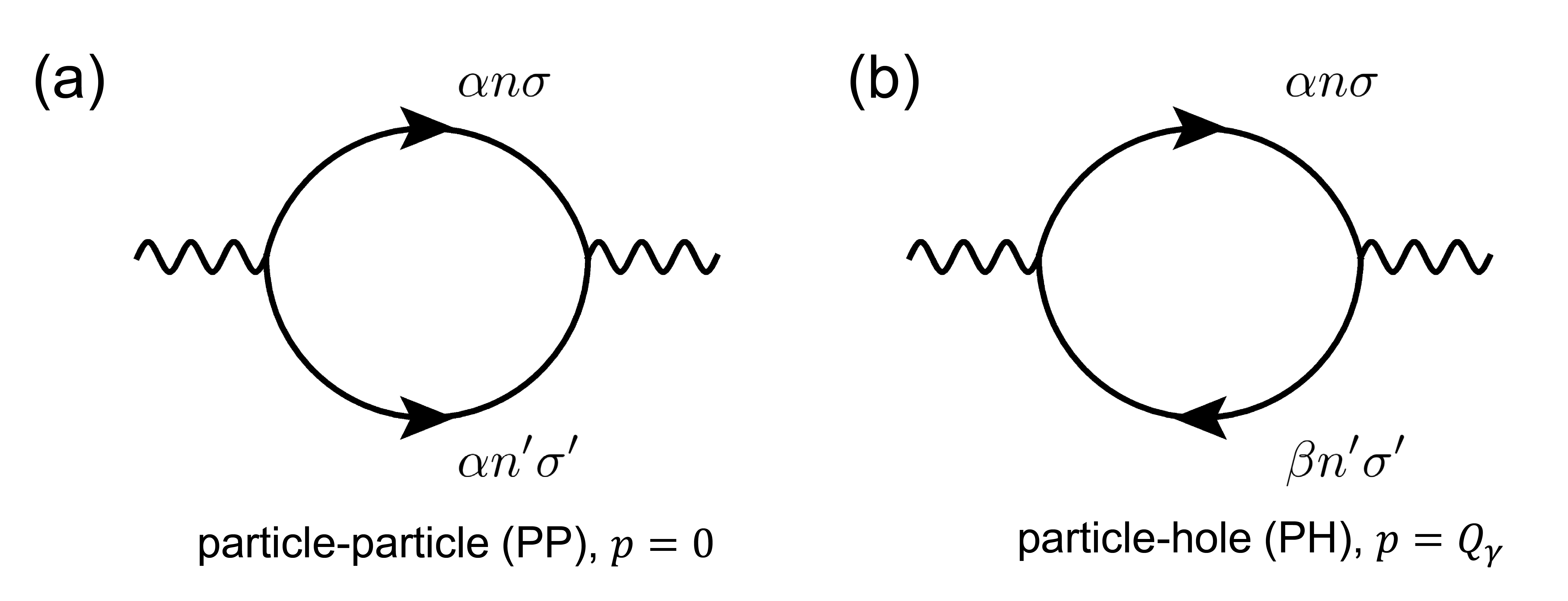}
\caption{Two diagrams that most divergent (log-squared) diagrams in one-loop perturbation. (a) Particle-particle susceptibility with $\bm q=0$. (b) Particle-hole susceptibility with $\bm q = \bm Q_\gamma$.
}
\label{fig:suscep}
\end{figure}

Due to the presence of VHSs, additional logarithmic divergences arise. Both the particle-particle channel at zero momentum (Cooper channel) and the particle-hole channel at finite momentum $\bm q = \bm Q_\alpha$ exhibit squared logarithmic divergences as follows~\cite{Furukawa1998,Nandkishore2012}:
\begin{eqnarray}
    \Pi^{\rm pp}(\bm q=0) &=& -T \sum_{i\omega} \int_{\bm k} \ \mathcal G_\alpha(i\omega, \bm k) \mathcal G_\alpha(-i\omega, -\bm k) \nn
    &\sim& \frac{h}{2} \log \frac{\Lambda}{{\rm max} (T,\mu)} \log \frac{\Lambda}{T}, \nn
    \Pi^{\rm ph}(\bm Q_\alpha) &=& T \sum_{i\omega} \int_{\bm k} \ \mathcal G_\beta(i\omega, \bm k) \mathcal G_\gamma(i\omega, \bm k + \bm Q_\alpha) \nn
    &\sim& h \log^2 \frac{\Lambda}{{\rm max}(T,\mu)}.
\end{eqnarray}
Here, $\mathcal G_\alpha(i\omega, \bm k) = (i\omega - \epsilon_\alpha(\bm k) + \mu)^{-1}$ is the Green's function of free part, $\mu$ is the chemical potential, $T$ is the temperature, $\Lambda$ is the ultraviolet cutoff, and $h = 1/(2\sqrt{3} \pi^2 t)$. The $p$RG method assumes these two channels dominate the RG flow while neglecting subleading divergent channels. Furthermore, it is convenient to use $y = \Pi^{\rm pp}(\bm q=0)$ as the RG flow parameter~\cite{Nandkishore2012,Lin2019,Lin2020}.

Within the one-loop approximation, the beta functions take the form
\begin{eqnarray}
    \beta_i(g) \equiv \frac{dg_i}{dy} = d_1 F_i^{\rm ph}(g) + F_i^{\rm pp}(g),
\label{eq:beta_func}
\end{eqnarray}
where $F_i^{\rm ph}(g)$ and $F_i^{\rm pp}(g)$ are quadratic polynomials of the coupling constants $g_i$. The five Feynman diagrams contributing to the beta functions are shown in Fig.~\ref{fig:RGdiag}, and the explicit forms of $F_i^{\rm ph/pp}$ are provided in Appendix~\ref{app:beta}.

The particle-hole channel involves a finite nesting momentum $\bm{Q}_\alpha$, and its RG flow is proportional to the nesting parameter $d_1(y) = d\Pi^{\rm ph}(\bm{Q}_\alpha)/dy$. This parameter ranges from $0$ to $0.5$, where the maximum value $d_1 = 0.5$ originates from the lack of perfect nesting in the kagome lattice. For simplicity, $d_1$ is treated as a constant with a value of $d_1 = 0.5$, as its functional dependence on $y$ has minimal impact on the results.

\begin{figure}[t!]
\includegraphics[width=\linewidth]{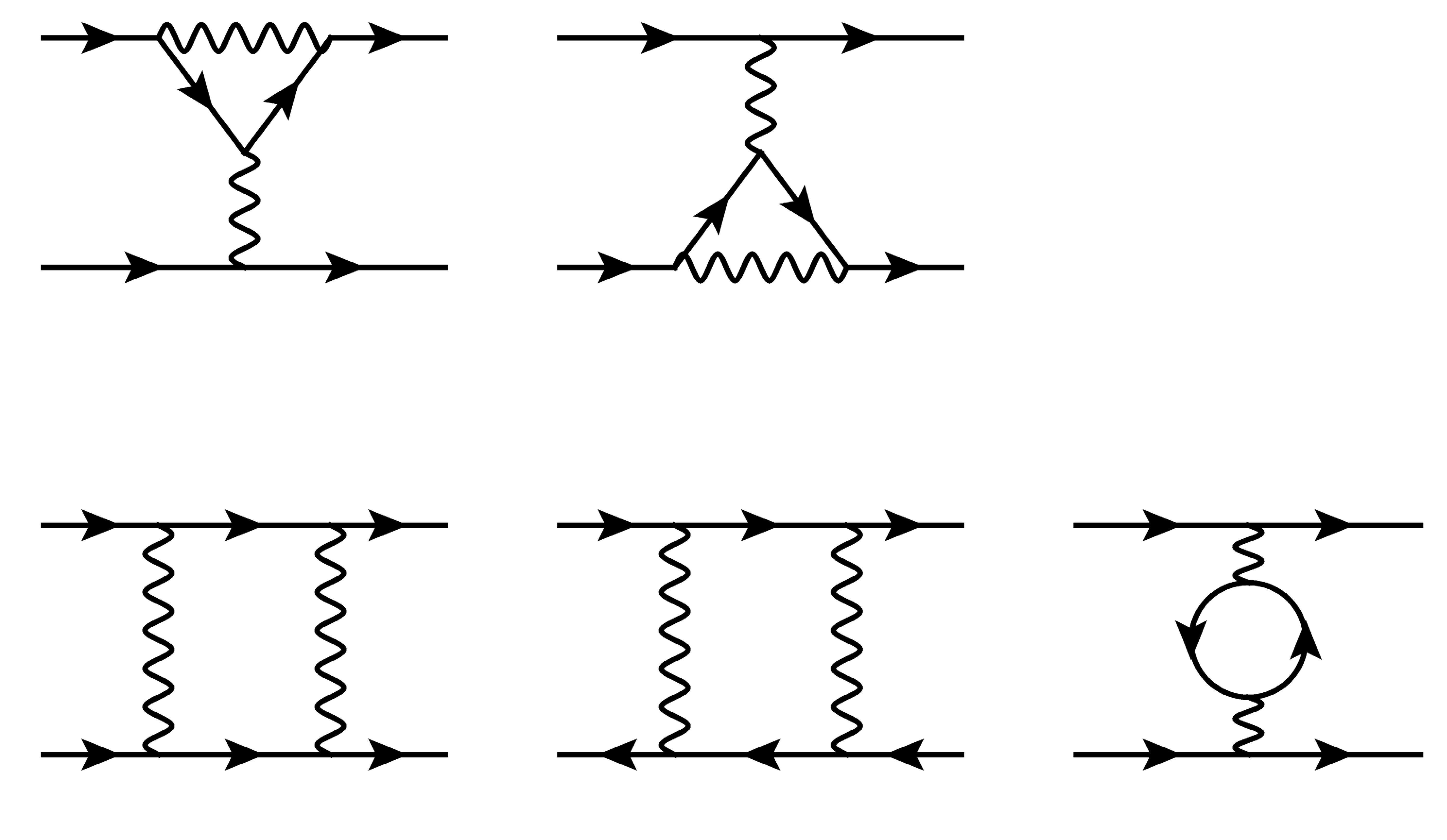}
\caption{Five one-loop Feynman diagrams with log-squared divergences contributing to the renormalization group flow. Note that the first and third diagrams in the second row yield negative contributions due to particle-particle pairing and the Fermion loop, respectively.
}
\label{fig:RGdiag}
\end{figure}

The explicit form of the beta functions reveals that $\beta_{22}$ and $\beta_{24}$ are positive semidefinite, while $\beta_{42}$ and $\beta_{44}$ are negative semidefinite. As a result, the couplings $g_{22}$, $g_{24}$, $g_{42}$, and $g_{44}$ exhibit runaway flow, and no fixed point exists other than $g_i = 0$. To address this, we adopt the fixed trajectory approach~\cite{Furukawa1998a,Furukawa1998,Binz2002,Nandkishore2012,Lin2019,Lin2020,Scammell2023}, as described below.

Since the beta functions are homogeneous polynomials of second order in $g_i$, the asymptotic behavior of the couplings is given by
\begin{eqnarray}
    g_i(y) \propto \frac{1}{y_c - y},
\end{eqnarray}
where $y_c$ is a critical value. Introducing new variables $G_i(y)$ defined as
\begin{eqnarray}
    G_i(y) = g_i(y) (y_c - y),
\label{eqn:gi_form}
\end{eqnarray}
and substituting this into Eq.~(\ref{eq:beta_func}), the beta functions for $G_i$ are obtained as
\begin{eqnarray}
    \frac{dG_i}{dy} = \frac{1}{y_c - y} \big[ \beta_i(G) - G_i \big].
\end{eqnarray}

The fixed trajectories correspond to the fixed points of this equation. To analyze the stability of the fixed point $G = G^*$, or equivalently the fixed trajectory, we linearize the equation to define the stability matrix,
\begin{eqnarray}
    M_{ij} = \frac{\partial (\beta_i(G) - G_i)}{\partial G_j} \Big|_{G=G^*}.
\end{eqnarray}
The stability is determined by the signs of the real parts of the eigenvalues of this matrix. For a single fixed trajectory to exist, the fixed point $G=G^*$ must be unstable in one direction and stable in all other directions. Although instability in more than one direction could be considered, all other trajectories are convergent when compared to the most divergent trajectories. Therefore, we only require solutions with one positive real part.

By solving the equation, we find four stable fixed trajectories, each determined by the initial values of the bare interactions. The critical interactions along these four stable trajectories are summarized in Table~\ref{tab:fix_traj}.

\begin{table}[b]
\caption{\label{tab:fix_traj}
The critical interactions along the four stable fixed trajectories.}
\begin{ruledtabular}
    \begin{tabular}{c|cccc}
        Channels & traj. \#1 & traj. \#2 & traj. \#3 & traj. \#4 \\
    \hline
    $g_{11}$ & -0.5  & -0.09050 & -0.01733 & 0.00001 \\
    $g_{12}$ & 0     & 0        & 0        & -0.00827 \\
    $g_{13}$ & -0.5  & 0        & 0        & -0.00310 \\
    $g_{14}$ & 0     & -0.04681 & 0.00991  & -0.00111 \\ 
    \hline
    $g_{21}$ & 0     & 0        & 0        & 0.00009 \\
    $g_{22}$ & 0     & 0.00755  & 0.00679  & 0.00531 \\
    $g_{23}$ & 0     & 0        & 0        & -0.00314 \\
    $g_{24}$ & 0     & 0.02702  & 0.02033  & 0.03258 \\
    \hline
    $g_{31}$ & 0     & 0.23090  & -0.20063 & -0.00036 \\
    $g_{32}$ & 0     & -0.12263 & -0.11633 & -0.03191 \\
    $g_{33}$ & 0     & 0        & 0        & 0.09781 \\
    $g_{34}$ & 0     & 0        & 0        & -0.25316 \\
    \hline
    $g_{41}$ & 0     & 0.23504  & -0.35355 & 0.00005 \\
    $g_{42}$ & 0     & -0.25906 & -0.36797 & -0.00204 \\
    $g_{43}$ & 0     & 0        & 0        & 0.19439 \\
    $g_{44}$ & 0     & 0        & 0        & -0.24525
    \end{tabular}
\end{ruledtabular}
\end{table}

\subsection{Instabilities}

To identify the instabilities of the system, we introduce test vertices and their corresponding susceptibilities. Here, we briefly outline the process, with detailed calculations provided in Appendix \ref{app:test}. The test vertices are particle-particle or particle-hole bilinears of the form,
\begin{eqnarray}
    \delta H = \sum \Delta c^\dagger c^{(\dagger)} + \text{h.c.}
\end{eqnarray}

The interactions $g_i$ induce corrections to the vertex, causing the test vertex $\Delta$ to evolve as the RG time $y$ increases. The flow equation for $\Delta$ is given by
\begin{eqnarray}
    \frac{d\Delta}{dy} = -g_\Delta \Delta,
\label{eqn:delta}
\end{eqnarray}
where $g_\Delta$ is a suitable linear combination of $g_i$. Using the form of Eq.~(\ref{eqn:gi_form}), $g_\Delta \approx G_\Delta / (y_c - y)$, the behavior of $\Delta$ is
\begin{eqnarray}
    \Delta \approx (y_c - y)^{G_\Delta}.
\end{eqnarray}

The instability manifests in the divergence of the susceptibility. After integrating out high-energy modes, including the test vertex, a term involving the susceptibility $\chi_\Delta$ remains in the Hamiltonian
\begin{eqnarray}
    \delta H = \sum \Delta^\dagger \chi_\Delta(y) \Delta.
\end{eqnarray}
The susceptibility $\chi_\Delta$ evolves with RG time $y$ as
\begin{eqnarray}
    \frac{d\chi_\Delta}{dy} = |\Delta|^2.
\end{eqnarray}
Near $y_c$, the susceptibility exhibits power-law behavior
\begin{eqnarray}
    \chi_\Delta \approx (y_c - y)^{\alpha_\Delta}.
\end{eqnarray}

The susceptibility exponent $\alpha_\Delta$ is related to the exponent of the test vertex as
\begin{eqnarray}
    \alpha_\Delta = 2G_\Delta + 1.
\end{eqnarray}
The sign of $\alpha_\Delta$ determines whether an instability occurs.

\begin{table*}[ht]
\caption{\label{tab:channels}
Irreducible channels, order parameters, and coupling constants derived from the RG flow of test vertices. $\lambda^a$ ($a = 1, 2, \cdots, 8$), $\tau^b$ ($b = 1, 2, 3$), and $\sigma^c$ ($c = 1, 2, 3$) denote the Gell-Mann matrices in patch space and the Pauli matrices in orbital and spin spaces, respectively. $[\lambda_\alpha^{|\epsilon|}]_{\beta\gamma} = |\epsilon^{\alpha\beta\gamma}|$ and $[\lambda_\alpha^{\epsilon}]_{\beta\gamma} = \epsilon^{\alpha\beta\gamma}$ represent components involving the Levi-Civita tensor $\epsilon^{\alpha\beta\gamma}$. Note that fixed trajectory \#1 leads to two degenerate instabilities, $r$CDW$_s^{\rm inter}$ and $i$CDW$_s^{\rm inter}$. For superconducting order parameters, the parities in spin space (even or odd) are explicitly indicated.}
\begin{ruledtabular}
    \begin{tabular}{cccccccc}
        Channel & Order parameter & Coupling constant & Leading \\
    \hline
        $r$CDW$^{\rm intra}_s$ & $N^{\rm intra}_{s,\alpha} = \frac{1}{2\sqrt 2} \sum_q \big< c_q^\dagger (\lambda_\alpha^{|\epsilon|} \otimes \tau^0 \otimes \sigma^0) c_q \big>$ & $g_{Ns} = g_{24}-2g_{14}-g_{34}+g_{21}-2g_{12}+g_{31}-2g_{32}$ & \\
        $i$CDW$^{\rm intra}_s$ & $\phi^{\rm intra}_{s,\alpha} = \frac{1}{2i\sqrt 2} \sum_q \big< c_q^\dagger (\lambda_\alpha^{\epsilon} \otimes \tau^0 \otimes \sigma^0) c_q \big>$ & $g_{\phi s} = g_{24}-2g_{14}+g_{34}+g_{21}-2g_{12}-g_{31}+2g_{32}$ & \\
        $r$SDW$^{\rm intra}_s$ & $\bm S^{\rm intra}_{s,\alpha} = \frac{1}{2\sqrt 2} \sum_q \big< c_q^\dagger (\lambda_\alpha^{|\epsilon|} \otimes \tau^0 \otimes \bm\sigma) c_q \big>$ & $g_{\bm S s} = g_{24} + g_{34} + g_{21} + g_{31}$ \\
        $i$SDW$^{\rm intra}_s$ & $\bm\psi^{\rm intra}_{s,\alpha} = \frac{1}{2i\sqrt 2} \sum_q \big< c_q^\dagger (\lambda_\alpha^{\epsilon} \otimes \tau^0 \otimes \bm\sigma) c_q \big>$ & $g_{\bm\psi s} = g_{24} - g_{34} + g_{21} - g_{31}$ \\
        $r$CDW$^{\rm intra}_a$ & $N^{\rm intra}_{a,\alpha} = \frac{1}{2\sqrt 2} \big< c_q^\dagger (\lambda_\alpha^{|\epsilon|} \otimes \tau^3 \otimes \sigma^0) c_q \big>$ & $g_{Na} = g_{24} - 2g_{14} - g_{34} - g_{21} + 2g_{12} - g_{31} + 2g_{32}$ \\
        $i$CDW$^{\rm intra}_a$ & $\phi^{\rm intra}_{a,\alpha} = \frac{1}{2i\sqrt 2} \big< c_q^\dagger (\lambda_\alpha^\epsilon \otimes \tau^3 \otimes \sigma^0) c_q \big>$, & $g_{\phi a} = g_{24} - 2g_{14} + g_{34} - g_{21} + 2g_{12} + g_{31} - 2g_{32}$ \\
        $r$SDW$^{\rm intra}_a$ & $\bm S^{\rm intra}_{a,\alpha} = \frac{1}{2\sqrt 2} \big< c_q^\dagger (\lambda_\alpha^{|\epsilon|} \otimes \tau^3 \otimes \bm\sigma) c_q \big>$ & $g_{\bm Sa} = g_{24} + g_{34} - g_{21} - g_{31}$ \\
        $i$SDW$^{\rm intra}_a$ & $\bm\psi^{\rm intra}_{a,\alpha} = \frac{1}{2i\sqrt 2} \big< c_q^\dagger (\lambda_\alpha^{\epsilon} \otimes \tau^3 \otimes \bm\sigma) c_q \big>$, & $g_{\bm\psi a} = g_{24} - g_{34} - g_{21} + g_{31}$ \\ \hline
        $r$CDW$^{\rm inter}_s$ & $N^{\rm inter}_{s,\alpha} = \frac{1}{2\sqrt 2} \sum_q \big< c_q^\dagger (\lambda_\alpha^{|\epsilon|} \otimes \tau^1 \otimes \sigma^0) c_q \big>$ & $\tilde g_{N s} = g_{24}-2g_{14}-g_{34}+g_{21}-2g_{12}+g_{31}-2g_{32}$ & \checkmark \\
        $i$CDW$^{\rm inter}_s$ & $\phi^{\rm inter}_{s,\alpha} = \frac{1}{2i\sqrt 2} \sum_q \big< c_q^\dagger (\lambda_\alpha^{\epsilon} \otimes \tau^1 \otimes \sigma^0) c_q \big>$ & $\tilde g_{\phi s} = g_{24}-2g_{14}+g_{34}+g_{21}-2g_{12}-g_{31}+2g_{32}$ & \checkmark \\
        $r$SDW$^{\rm inter}_s$ & $\bm S^{\rm inter}_{s,\alpha} = \frac{1}{2\sqrt 2} \sum_q \big< c_q^\dagger (\lambda_\alpha^{|\epsilon|} \otimes \tau^1 \otimes \bm\sigma) c_q \big>$ & $\tilde g_{\bm S s} = g_{24} + g_{34} + g_{21} + g_{31}$ \\
        $i$SDW$^{\rm inter}_s$ & $\bm\psi^{\rm inter}_{s,\alpha} = \frac{1}{2i\sqrt 2} \sum_q \big< c_q^\dagger (\lambda_\alpha^{\epsilon} \otimes \tau^1 \otimes \bm\sigma) c_q \big>$ & $\tilde g_{\bm\psi s} = g_{24} - g_{34} + g_{21} - g_{31}$ \\
        $r$CDW$^{\rm inter}_a$ & $N^{\rm inter}_{a,\alpha} = \frac{1}{2\sqrt 2} \big< c_q^\dagger (\lambda_\alpha^{|\epsilon|} \otimes i\tau^2 \otimes \sigma^0) c_q \big>$ & $\tilde g_{N a} = g_{24} - 2g_{14} - g_{34} - g_{21} + 2g_{12} - g_{31} + 2g_{32}$ \\
        $i$CDW$^{\rm inter}_a$ & $\phi^{\rm inter}_{a,\alpha} = \frac{1}{2i\sqrt 2} \big< c_q^\dagger (\lambda_\alpha^\epsilon \otimes i\tau^2 \otimes \sigma^0) c_q \big>$, & $\tilde g_{\phi a} = g_{24} - 2g_{14} + g_{34} - g_{21} + 2g_{12} + g_{31} - 2g_{32}$ \\
        $r$SDW$^{\rm inter}_a$ & $\bm S^{\rm inter}_{a,\alpha} = \frac{1}{2\sqrt 2} \big< c_q^\dagger (\lambda_\alpha^{|\epsilon|} \otimes \tau^3 \otimes \bm\sigma) c_q \big>$ & $\tilde g_{\bm S a} = g_{24} + g_{34} - g_{21} - g_{31}$ \\
        $i$SDW$^{\rm inter}_a$ & $\bm\psi^{\rm inter}_{a,\alpha} = \frac{1}{2i\sqrt 2} \big< c_q^\dagger (\lambda_\alpha^{\epsilon} \otimes i\tau^2 \otimes \bm\sigma) c_q \big>$, & $\tilde g_{\bm\psi a} = g_{24} - g_{34} - g_{21} + g_{31}$ \\ \hline
        SC$^{\rm intra}_{sOS}$ (even) & $\Delta_{sOS} = \frac{1}{\sqrt 6} \sum_{q} \big< c_{q}^\dagger (\lambda^0 \otimes \tau^0 \otimes i\sigma^2) c_{-q}^\dagger \big>$ & $g_{sOS} = g_{44} + 2g_{34} + g_{43} + 2g_{33}$ & \\
        SC$^{\rm intra}_{sZS}$ (even) & $\Delta_{sZS} = \frac{1}{\sqrt 6} \sum_q \big< c_{q}^\dagger (\lambda^0 \otimes \tau^3 \otimes i\sigma^2) c_{-q}^\dagger \big>$ & $g_{sZS} = g_{44} + 2g_{34} - g_{43} - 2g_{33}$ & \checkmark \\
        SC$^{\rm intra}_{dOS}$ (even) & $ \Delta_{dOS} = \frac{1}{2} \sum_{q} \big< c_{q}^\dagger (\lambda^3 \otimes \tau^0 \otimes i\sigma^2) c_{-q}^\dagger \big>$ & $g_{dOS} = g_{44} - g_{34} + g_{43} - g_{33}$ & \\
        SC$^{\rm intra}_{\bar dOS}$ (even) & $\Delta_{\bar dOS} = \frac{1}{2} \sum_q \big< c_q^\dagger (\lambda^8 \otimes \tau^0 \otimes i\sigma^2) c_{-q}^\dagger \big>$ & $g_{\bar dOS} = g_{44} - g_{34} + g_{43} - g_{33}$ & \\
        SC$^{\rm intra}_{dZS}$ (even) & $\Delta_{dZS} = \frac{1}{2} \sum_{q} \big< c_{q}^\dagger (\lambda^3 \otimes \tau^3 \otimes i\sigma^2) c_{-q}^\dagger \big>$ & $g_{dZS} = g_{44} - g_{34} - g_{43} + g_{33}$ & \\
        SC$^{\rm intra}_{\bar d ZS}$ (even) & $\Delta_{\bar dZS} = \frac{1}{2} \sum_q \big< c_q^\dagger (\lambda^8 \otimes \tau^3 \otimes i\sigma^2) c_{-q}^\dagger \big>$ & $g_{\bar dZS} = g_{44} - g_{34} - g_{43} + g_{33}$ \\ \hline
        SC$^{\rm inter}_{sXS}$ (even) & $\Delta_{sXS} = \frac{1}{\sqrt 6} \sum_{q} \big< c_{q}^\dagger (\lambda^0 \otimes \tau^1 \otimes i\sigma^2) c_{-q}^\dagger \big>$ & $g_{sXS} = g_{42} + 2g_{32} + g_{41} + 2g_{31}$ & \checkmark \\
        SC$^{\rm inter}_{siYT}$ (odd) & $\bm\Delta_{siYT} = \frac{1}{\sqrt 6} \sum_{q} \big< c_{q}^\dagger (\lambda^0 \otimes i\tau^2 \otimes \bm\sigma i\sigma^2) c_{-q}^\dagger \big>$ & $g_{siYT} = g_{42} + 2g_{32} - g_{41} - 2g_{31}$ & \checkmark \\
        SC$^{\rm inter}_{dXS}$ (even) & $\Delta_{dXS} = \frac{1}{2} \sum_{q} \big< c_{q}^\dagger (\lambda^3 \otimes \tau^1 \otimes i\sigma^2) c_{-q}^\dagger \big>$ & $g_{dXS} = g_{42} - g_{32} + g_{41} - g_{31}$ \nn
        SC$^{\rm inter}_{\bar dXS}$ (even) & $\Delta_{\bar dXS} = \frac{1}{2} \sum_q \big< c_q^\dagger (\lambda^8 \otimes \tau^1 \otimes i\sigma^2) c_{-q}^\dagger \big>$ & $g_{\bar dXS} = g_{42} - g_{32} + g_{41} - g_{31}$ \nn
        SC$^{\rm inter}_{diYT}$ (odd) & $\bm \Delta_{diYT} = \frac{1}{2} \sum_{q} \big< c_{q}^\dagger (\lambda^3 \otimes i\tau^2 \otimes \bm\sigma i\sigma^2) c_{-q}^\dagger \big>$ & $g_{diYT} = g_{42} - g_{32} - g_{41} + g_{31}$ \nn
        SC$^{\rm inter}_{\bar diYT}$ (odd) & $\bm\Delta_{\bar diYT} = \frac{1}{2} \sum_q \big< c_q^\dagger (\lambda^8 \otimes i\tau^2 \otimes \bm\sigma i\sigma^2) c_{-q}^\dagger \big>$ & $g_{\bar diYT} = g_{42} - g_{32} - g_{41} + g_{31}$
    \end{tabular}
\end{ruledtabular}
\end{table*}

\section{Results}

\subsection{Test vertices}

The test vertices we consider are defined as follows. For the particle-hole or density-wave channels, the perturbation is given by
\begin{eqnarray}
    \delta H = \sum_{\alpha \neq \beta, n, n', \sigma, \sigma'} \Phi_{\gamma, n \sigma, n' \sigma'} c_{\beta n \sigma}^\dagger c_{\alpha n' \sigma'} + \text{h.c.}
\label{eq:HamilDW}
\end{eqnarray}
For the particle-particle or superconducting pairing channels, it is
\begin{eqnarray}
    \delta H = \sum_{\alpha, n, n', \sigma, \sigma'} \Delta_{\alpha, n \sigma, n' \sigma'} c_{\alpha n \sigma}^\dagger c_{\alpha n' \sigma'}^\dagger + \text{h.c.}
\label{eq:HamilSC}
\end{eqnarray}
Here, $\Phi$ and $\Delta$ are the density-wave and superconducting order parameters, respectively. Note that $\Phi$ carries a momentum $\bm{q} = \bm{Q}_\gamma$, while $\Delta$ does not.

\begin{figure}[t!]
\includegraphics[width=0.6\linewidth]{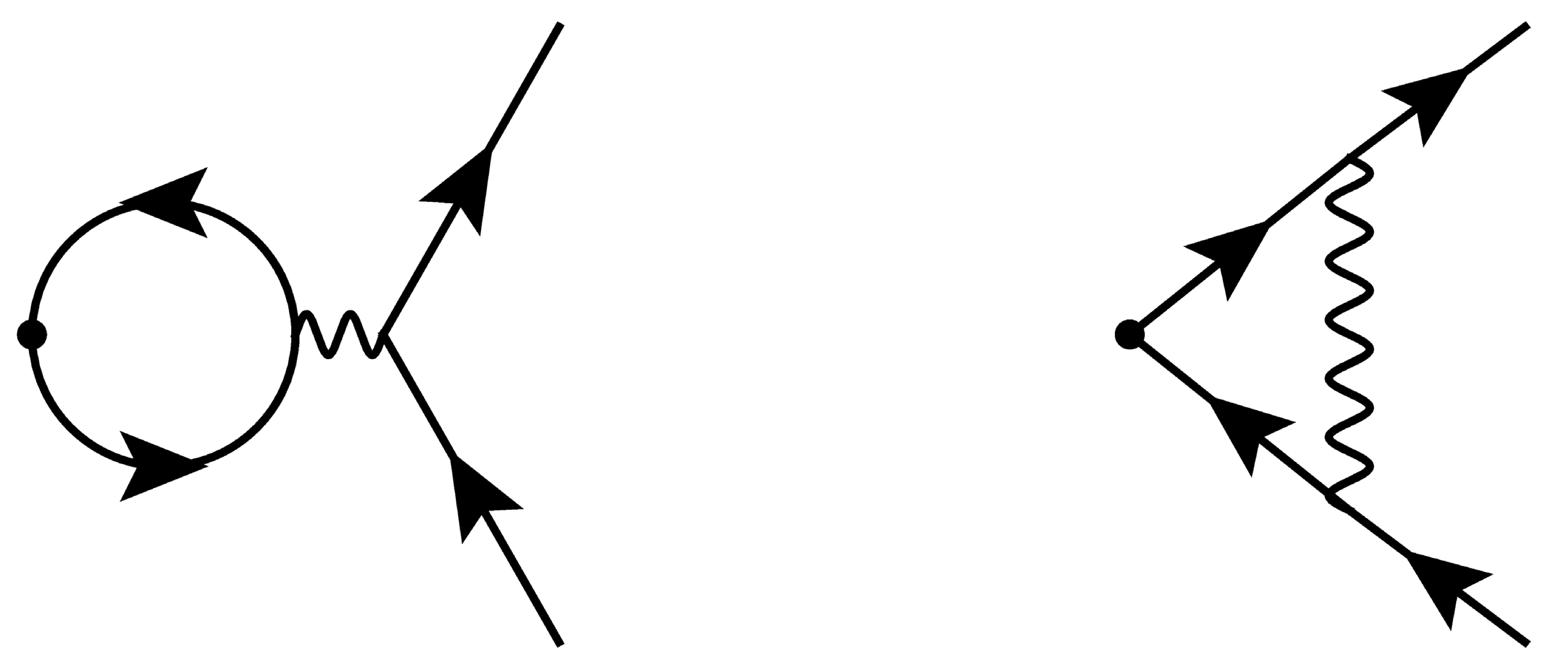}
\caption{Two Feynman diagrams for vertex corrections in one-loop order.
}
\label{fig:RGtest}
\end{figure}

The one-loop renormalization of the test vertices, or vertex corrections, is represented by the diagrams shown in Fig.~\ref{fig:RGtest}. At the one-loop level, the corrections are linear in the coupling constants $g_i$. For density channels, the corrections are described by the following flow equation,
\begin{eqnarray}
    \frac{d\Phi}{dy} = d_1 A_{\Phi}(g) \Phi.
\label{eq:test_flow}
\end{eqnarray}
Here, $\Phi$ is a 48-dimensional vector, where the dimensionality arises from the combination of three patch indices and four orbital-spin indices for both incoming and outgoing particles ($3 \times 4^2 = 48$). Thus, $A_\Phi$ is $48\times48$ matrix whose components are linear in $g_i$. The detailed forms of these matrices are provided in Appendix~\ref{app:test}.

By diagonalizing the $A_\Phi$ matrix, the irreducible representations (irreps) of the order parameters $\Phi$ under the lattice symmetry can be determined. Irreps are categorized based on the $\bm Q_\alpha$ vectors connecting M-points (3), their charge or spin character (4), their inter- or intra-orbital nature (2), and whether they form symmetric or antisymmetric configurations (2), resulting in a total of 48 (=$3\times4\times2\times2$) irreps. Note that the absence of spin-orbit coupling in our model leads to degeneracy in the three spin directions. Table~\ref{tab:channels} presents all irreps, further distinguishing between the real and imaginary parts of the order parameters. These irreps are denoted as, for example, $r\text{CDW}_s^{\text{intra}}$ for intra-orbital symmetric real charge density wave (CDW) order, and $i\text{SDW}_a^{\text{inter}}$ for inter-orbital antisymmetric imaginary spin density wave (SDW) order. The detailed procedure for deriving these irreps described in Appendix~\ref{app:test}.

For pairing channels, the corrections are described by the following flow equation,
\begin{eqnarray}
    \frac{d\Delta}{dy} = A_{\Delta}(g) \Delta.
\label{eq:test_flow}
\end{eqnarray}
Here, $\Delta$ is 18-dimensional vector due to the antisymmetric condition for the pairing fields. Similar to the CDW case, one can divide the space into irreps by diagonalizing $A_\Delta$. The irrpes are categorized into angular momentum symmetry in patch space ($s$- or $d$-wave, 3), intra-orbital spin-singlet configurations (2), and inter-orbital spin-singlet or spin-triplet configurations (4), resulting in a total of 18 (=$3 \times (2+4)$) irreps. The full list of irreps, order parameters, and their corresponding coupling constants is provided in Table~\ref{tab:channels}, with the detailed procedure for deriving these irreps described in Appendix~\ref{app:test}. We label the pairing order parameters in terms of their structure in patch, orbital, and spin spaces as 
\begin{equation}
    \Delta_{\alpha n \sigma} = \big\langle c_{-k} \mathcal{P}_\alpha \otimes \mathcal{O}_n \otimes \mathcal{S}_\sigma \, c_k \big\rangle.
\end{equation}
Here, the patch index $\alpha = s, d, \bar{d}$ corresponds to $\mathcal{P}_\alpha = \lambda^0 = \sqrt{\frac{2}{3}} \mathds{1}_3$, $\lambda^3$, $\lambda^8$, where $\lambda^a$ are Gell-Mann matrices. The orbital indices are $n = O, Z, X, iY$, corresponding to $\mathcal{O}_n = \tau^0$, $\tau^3$, $\tau^1$, and $i\tau^2$, where $\tau^b$ are Pauli matrices in orbital space. The spin index $\sigma = S, T$ corresponds to $\mathcal{S}_\sigma = i\sigma^2$, $\bm{\sigma} i\sigma^2$, where $\sigma^c$ are Pauli matrices in spin space. For example, the pairing $\Delta_{diYT}$ is defined as
\begin{equation}
    \Delta_{diYT} = \big\langle c_{-k} \lambda^3 \otimes i\tau^2 \otimes \bm{\sigma} i\sigma^2 \, c_k \big\rangle.
\end{equation}

Since density-wave instabilities have been extensively studied but in other contexts~\cite{Scammell2023}, our focus here is on superconducting instabilities. These are particularly intriguing due to their potential to unveil unconventional pairing mechanisms and their sensitivity to the underlying lattice symmetries. By examining the RG flow of the couplings and the structure of irreps, we identify the dominant superconducting order parameters and discuss their physical implications in the following sections.

\subsection{Phase diagrams}

\begin{figure}[t!]
\includegraphics[width=\linewidth]{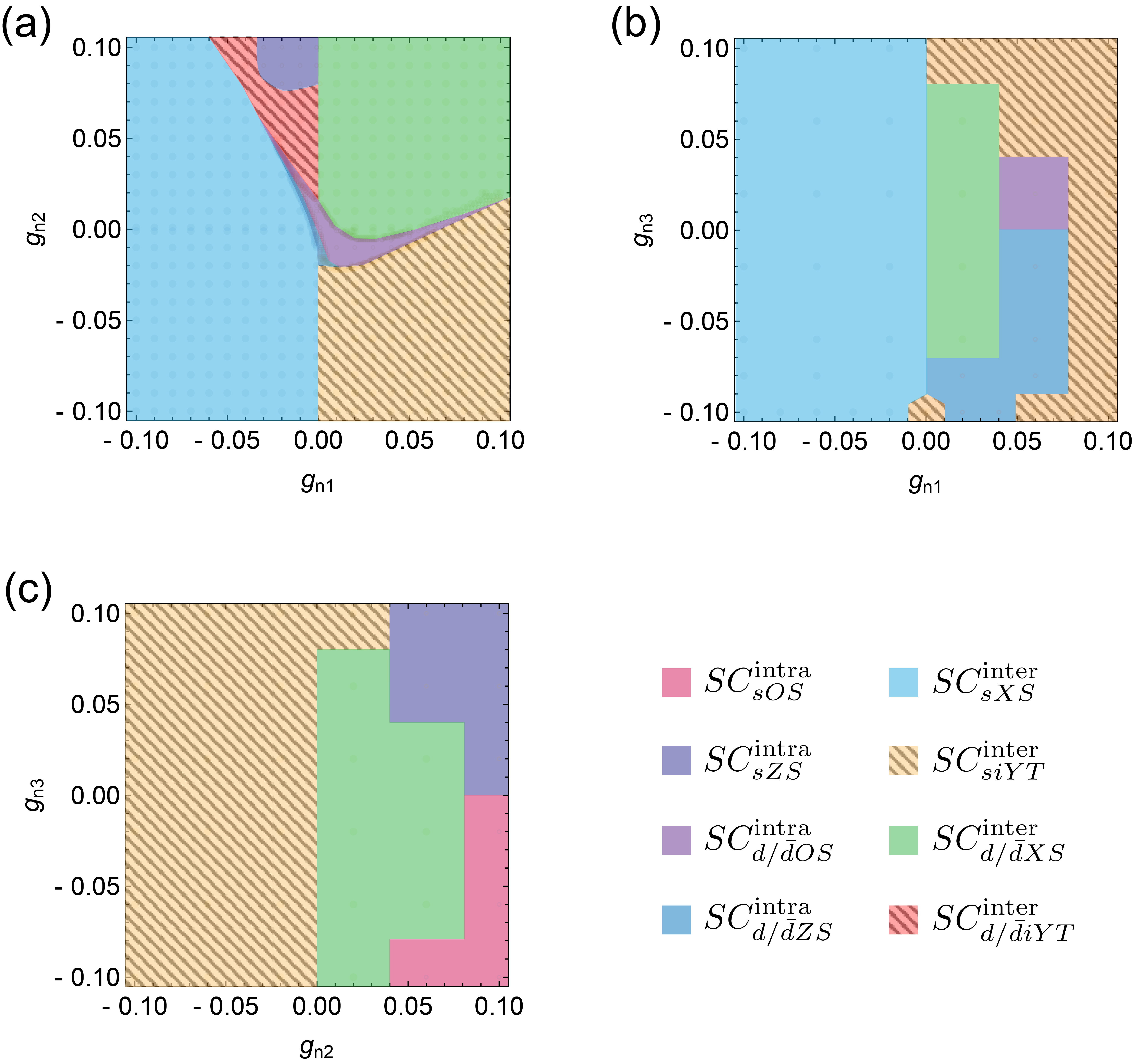}
\caption{Phase diagrams of superconducting order parameters as functions of the intra- and inter-orbital interaction parameters,
(a) $g_1$ versus $g_2$, (b) $g_1$ versus $g_3$, and (c) $g_2$ versus $g_3$. 
The intra-coupling values are fixed as $g_{14} = g_{24} = g_{34} = 0.1$ and $g_{44} = 0.3$, assuming all $g_{ab}$ with $b \neq 4$ are equal for simplicity. 
The diagrams reveal parameter regions where spin-triplet superconducting phases, such as SC$^{\rm inter}_{sXT}$ and SC$^{\rm inter}_{diYT}$, emerge as leading instabilities.
}
\label{fig:phase}
\end{figure}

Based on the RG analysis presented in the previous section, we construct the phase diagram shown in Fig.~\ref{fig:phase}. The phases are determined by identifying the most negative susceptibility exponent $\alpha_\Delta$ for various initial (bare or UV) interaction values in the RG flow. Due to the large number of coupling constant combinations (120 pairs), we restrict our search to simplified subspaces. Specifically, we assume that all $g_{ab}$ with $b \neq 4$ are equal, neglecting differences between intra-orbital processes for a given inter-orbital channel.

We note that there are parameter regions where spin-triplet superconductivity emerges as the leading instability (denoted by the orange and red regions with a lined pattern). The phase in the lined-orange region corresponds to the SC$^{\rm inter}_{sXT}$ phase, while the lined-red region corresponds to the SC$^{\rm inter}_{diYT}$ phase. The SC$^{\rm inter}_{sXT}$ phase arises when $g_{a1} > 0$ and $g_{a2} < 0$, whereas the SC$^{\rm inter}_{diYT}$ phase occurs for $g_{a1} < 0$ and $g_{a2} > 0$. Recall that $g_{a1}$ represents the exchange interaction and $g_{a2}$ represents the density-density interaction in the orbital space.

\section{Landau-Ginzberg functional analysis for degenerate phases}

As illustrated in Table~\ref{tab:channels}, we identified pairs of degenerate superconducting phases, namely \{SC$^{\rm intra}_{dOS}$, SC$^{\rm intra}_{\bar dOS}$\}, \{SC$^{\rm inter}_{dZS}$, SC$^{\rm inter}_{\bar dZS}$\} and \{SC$^{\rm inter}_{diYT}$, SC$^{\rm inter}_{\bar diYT}$\}. For intra-orbital phases, the results are consistent with those obtained for single-orbital models. In this work, we focus specifically on the inter-orbital pairs, \{SC$^{\rm inter}_{dZS}$, SC$^{\rm inter}_{\bar dZS}$\} and \{SC$^{\rm inter}_{diYT}$, SC$^{\rm inter}_{\bar diYT}$\}. To understand the phase of the ground state within this degenerate space, we employ a Landau-Ginzberg-type analysis. By considering the symmetry-allowed terms in the free energy, we analyze the interactions and couplings between degenerate order parameters to determine the most energetically favorable configuration of the system. Specifically, we identify whether the ground state adopts a chiral state where the time-reversal symmetry is broken or a helical state where the time-reversal symmetry is preserved.

\subsection{Degenerate inter-orbital spin-singlet SC phase}

Intra-orbital pairing phases are similar to the single orbital phases. The only difference is the splitting of the coupling constant due to the correlation between two orbitals. There can be symmetric and antisymmetric orbitals,
\begin{eqnarray}
    \Delta_{dO/ZS} = \frac{1}{2} \sum_q \big< c_{-q} (\lambda^3 \otimes \tau^{0/3} \otimes i\sigma^2) c_q \big> \equiv \Delta_1, \nn
    \Delta_{\bar d O/Z S} = \frac{1}{2} \sum_q \big< c_{-q} (\lambda^8 \otimes \tau^{0/3} \otimes i\sigma^2) c_q \big> \equiv \Delta_2.
\end{eqnarray}
Here, $\lambda^a$ ($a=0, 1, \cdots, 8$) are Gell-Mann matrices with identity in patch space, $\tau^b$ and $\sigma^b$ ($b=0, 1, 2, 3$) are Pauli matrices with identity in orbital and spin space, respectively. The Landau free-energy is
\begin{eqnarray}
    F = c_1 (|\Delta_1|^2 + |\Delta_2|^2) + c_4 (|\Delta_1|^2 + |\Delta_2|^2)^2 \nn
    - K |\Delta_1|^2 |\Delta_2|^2 \sin^2\phi,
\label{eq:LG1}
\end{eqnarray}
where $\phi = \arg \Delta_1/\Delta_2$. $c_1 \propto T-T_c$ determines phase transition temperautre $T_c$ and $c_4>0$. In the ordered phase, $T<T_c$, both order parameters $\Delta_1$ and $\Delta_2$ are present, and the feature of the ground state is determined by the sign of $K$. For positive $K$, the free energy is minimum for $\phi=0,\pi$ which yields 
\begin{eqnarray}
    \Delta^\pm = \Delta_1 \pm \Delta_2.
\end{eqnarray}
Thus, the phase preserves time-reversal symmetry.

For negative $K$, the order parameter favors $\phi = \pm \pi/2$, 
\begin{eqnarray}
    \Delta^\pm_{O/Z} = \Delta e^{\pm 2i\theta_k} \tau^{0/3} i\sigma^2,
\end{eqnarray}
where $\theta_k = 2\pi/3$. This is a chiral SC phase that breaks time-reversal symmetry. 

\subsection{Inter-orbital spin-triplet SC phase}

There are degenerate $d$-wave-like orbital-singlet spin-triplet SC phases. The order parameters are
\begin{eqnarray}
    \bm\Delta_{1} = \frac{1}{2} \sum_q \big< c_{-q} (\lambda^3 \otimes i\tau^2 \otimes \bm \sigma i\sigma^2) c_q \big>, \nn
    \bm\Delta_{2} = \frac{1}{2} \sum_q \big< c_{-q} (\lambda^8 \otimes i\tau^2 \otimes \bm \sigma i\sigma^2) c_q \big>.
\end{eqnarray}
The Landau free energy has a form
\begin{eqnarray}
    F &=& c_2 \big( |\bm\Delta_1|^2 + |\bm\Delta_2|^2 \big) + c_4 \bigg[ \big( |\bm\Delta_1|^2 + |\bm\Delta_2|^2 \big)^2 \nn
    &&+ |\bm\Delta_1^* \times \bm\Delta_1|^2 + |\bm\Delta_2^* \times \bm\Delta_2|^2 \nn
    &&+ \frac{1}{3} \bigg\{ -2|\bm\Delta_1|^2 |\bm\Delta_2|^2 + \bm\Delta_1^2 \bm\Delta_2^{*2} + \bm\Delta_1^{*2} \bm\Delta_2^2 \nn
    &&- 2\big( \bm\Delta_1 \times \bm\Delta_2^* \big)^2 - 2\big( \bm\Delta_1^* \times \bm\Delta_2 \big)^2 \nn
    &&- 4\big| \bm\Delta_1 \times \bm\Delta_2^* \big|^2 + 4\big| \bm\Delta_1 \times \bm\Delta_2 \big|^2 \bigg\} \bigg].
\label{eq:LG2}
\end{eqnarray}
The constant $c_2 \propto T-T_c$ and $c_4>0$ to develop superconducting phase transition at $T=T_c$. The ground state is determined by the anisotropic terms in quartic order. Define the angles $\theta = \cos^{-1} (\bm n_1 \cdot \bm n_2)$ and $\phi = \arg (\Delta_2 / \Delta_1)$ with $\bm\Delta_a = \Delta_a \bm n_a$, $|\bm n_a|=1$, $a=1,2$. Then, the free energy is
\begin{eqnarray}
    F &=& c_2 \big(\Delta_1^2 + \Delta_2^2\big) + c_4 \Big[ \big( \Delta_1^2 + \Delta_2^2 \big)^2 \nn
    &&-\frac{4}{3} \Delta_1^2 \Delta_2^2 \big( \cos^2\theta \sin^2\phi + \sin^2\theta \cos^2\phi \big) \Big].
\end{eqnarray}
From the second line, one can deduce the ground state order parameters. The first one is the $p+ip$ chiral order, $\theta=0,\pi$ and $\phi=\pm \pi/2$. Thus, the order parameter becomes $\bm\Delta_2 = \pm i\bm\Delta_1$ which breaks time-reversal symmetry. The second order parameter is helical order, $\theta=\pi/2$ and $\phi=0,\pi$. Thus, the order parameter is $\bm\Delta_2 = \pm \bm \Delta_1$, and the time-reversal symmetry is preserved.

\section{Conclusion}

In this study, we investigated the effects of multi-orbital interactions on superconducting phases in kagome metals using $p$RG method. There can be 18 possible pairing functions due to degrees of freedom in patch, orbital, and spin space. Through the $p$RG analysis, we identified four stable fixed trajectories. Phase diagrams are obtained in terms of bare coupling strengths, which reveal eight possible phases. Notably, there is a region of $d$-wave-like orbital-singlet spin-triplet phase. This phase allows time-reversal-symmetry broken spin-triplet phase or chiral spin-triplet phase.

Experimentally, kagome metals with multiple VHSs near the Fermi energy have been identified. Examples include vanadium-based materials $A$V$_3$Sb$_5$ ($A$ = K, Rb, Cs)~\cite{Jiang2021,Li2022,Liu2021,Frassineti2023,Song2023} and FeGe~\cite{Teng2022,Yin2022a}. The electronic structure of these materials can be modified by applying pressure~\cite{Chen2021a,Ni2021} or chemical substitution~\cite{Li2022a,Song2024,Yousuf2024}, allowing for the realization and control of multiple VHSs near the Fermi energy. This tunability provides a promising avenue for exploring the effects of the multiple VHS on the correlation-driven electronic phases in kagome metals.

Furthermore, the emergence of a new superconducting phase has been observed under pressure~\cite{Zhang2021} or chemical substitution~\cite{Sur2023}. This suggests a possible connection between multiple VHSs and superconductivity, where the enhanced density of states at the Fermi level could drive unconventional pairing mechanisms. Our theoretical framework can be applied to explain these phenomena and indicates that the resulting superconducting phase may exhibit a chiral spin-triplet nature. These findings highlight the importance of multiple VHSs in kagome metals and their potential role in stabilizing exotic superconducting states.

In conclusion, our work highlights the significance of the multi-orbital nature in the emergence of superconductivity in systems with VHSs, a concept that can be extended to other materials exhibiting VHSs beyond kagome metals. Additionally, future studies should explore the inclusion of spin-orbit coupling and electron-phonon interactions to provide a more comprehensive understanding of the complex physics in kagome metals and other multi-band systems.

\begin{acknowledgments}
J.H. and S.B.L. were supported by National Research Foundation Grant (2021R1A2C109306013). 
\end{acknowledgments}

\appendix

\section{Beta functions}
\label{app:beta}

The functions $F_i^{\rm ph}(g)$ and $F_i^{pp}(g)$ in the beta functions, Eq.~(\ref{eq:beta_func}), are given below:
\begin{widetext}
\begin{eqnarray}
    F_{11}^{\rm ph} &=& d \Big( 2g_{11} g_{22} + 2g_{13} g_{23} - N_f g_{11}^2 - N_f g_{13}^2 + 4g_{31} g_{32} + 2g_{33}^2 - N_f g_{31}^2 - N_f g_{33}^2 \Big), \ \ \ F_{11}^{\rm pp} = 0, \nn
    F_{12}^{\rm ph} &=& d \Big( 2g_{12} g_{24} + 2g_{14} g_{21} - 2N_f g_{12} g_{14} + 4g_{32} g_{34} + 4g_{34} g_{31} - 2N_f g_{32} g_{34} \Big), \ \ \ F_{12}^{\rm pp} = 0, \nn
    F_{13}^{\rm ph} &=& d \Big( 2g_{11} g_{23} + 2g_{13} g_{22} - 2N_f g_{11} g_{13} + 4g_{31} g_{33} + 4g_{33} g_{32} - 2N_f g_{31} g_{33} \Big), \ \ \ F^{\rm pp}_{13} = 0, \nn
    F_{14}^{\rm ph} &=& d \Big( 2g_{14} g_{24} - N_f g_{14}^2 + 2g_{44}^2 - N_f g_{34}^2 
    + 2g_{12} g_{21} - N_f g_{12}^2 + 4g_{32} g_{31} - N_f g_{32}^2 \Big), \ \ \ F^{\rm pp}_{14} = 0,
\end{eqnarray}
\begin{eqnarray}
    F_{21}^{\rm ph} &=& 2d \Big( g_{21} g_{24} + g_{31} g_{34} \Big), \ \ \ F^{\rm pp}_{21} = 0, \nn
    F_{22}^{\rm ph} &=& d \Big( g_{22}^2 + g_{32}^2 + g_{23}^2 + g_{33}^2 \Big), \ \ \ F^{\rm pp}_{22} = 0, \nn
    F_{23}^{\rm ph} &=& 2d \Big( g_{22} g_{23} + g_{32} g_{33} \Big), \ \ \ F^{\rm pp}_{23} = 0, \nn
    F_{24}^{\rm ph} &=& d \Big( g_{21}^2 + g_{31}^2 + g_{24}^2 + g_{34}^2 \Big), \ \ \ F^{\rm pp}_{24} = 0,
\end{eqnarray}
\begin{eqnarray}
    F_{31}^{\rm ph} &=& 2d \Big( g_{22} g_{31} + 2g_{23} g_{33} + g_{21} g_{34} - N_f g_{11} g_{31} - N_f g_{13} g_{33} + g_{11} g_{32} + g_{13} g_{31} + g_{13} g_{33} \Big) \nn
    && F^{\rm pp}_{31} = - 2p \Big[ (N_p-2) g_{31} g_{32} + g_{31} g_{42} \Big], \nn
    F_{32}^{\rm ph} &=& 2d \Big( - N_f g_{12} g_{34} + g_{24} g_{32} + g_{12} g_{34} + g_{14} g_{31} \Big) + 2\bar d g_{21} g_{34} + 2d_{int} g_{22} g_{32} \nn
    && F^{\rm pp}_{32} = - p_{int} \Big[ (N_p-2) g_{31}^2 + g_{31} g_{41} \Big] - \bar p \Big[ (N_p-2) g_{33}^2 + g_{33} g_{43} \Big], \nn
    F_{33}^{\rm ph} &=& 2d_{int} \Big( g_{23} g_{33} + g_{23} g_{31} + g_{22} g_{33} - N_f g_{11} g_{33} + g_{11} g_{33} + g_{13} g_{32} \Big) \nn
    && F^{\rm pp}_{33} = - 2p \Big[ (N_p-2) g_{33} g_{34} + g_{33} g_{44} \Big], \nn
    F_{34}^{\rm ph} &=& 2d\Big( 2g_{24} g_{34} - N_f g_{14} g_{34} + g_{14} g_{34} \Big) - 2N_f \bar d g_{12} g_{32} + 2d_{int} \Big( g_{21} g_{32} + g_{21} g_{31} + g_{12} g_{31} \Big) \nn
    && F^{\rm pp}_{34} = - p \Big[ (N_p-2) g_{34}^2 + 2g_{34} g_{44} \Big] - \bar p \Big[ (N_p-2) g_{33}^2 + 2g_{33} g_{43} \Big],
\end{eqnarray}
\begin{eqnarray}
    F_{41}^{\rm ph} &=& 0, \ \ \ 
    F^{\rm pp}_{41} = -2p_{int} \Big[ (N_p-1) g_{31} g_{32} + g_{41} g_{42} \Big], \nn
    F_{42}^{\rm ph} &=& 0, \ \ \ 
    F^{\rm pp}_{42} = - p_{int} \Big[ (N_p-1) g_{31}^2 + (N_p-1) g_{32}^2 + g_{41}^2 + g_{42}^2 \Big], \nn
    F_{43}^{\rm ph} &=& 0, \ \ \ 
    F^{\rm pp}_{43} = - p \Big[ (N_p-1) g_{33} g_{34} + g_{43} g_{44} \Big] - \bar p \Big[ (N_p-1) g_{33} g_{34} + g_{43} g_{44} \Big], \nn
    F_{44}^{\rm ph} &=& 0, \ \ \ 
    F^{\rm pp}_{44} = -p \Big[ (N_p-1) g_{34}^2 + g_{44}^2 \Big] - \bar p \Big[ (N_p-1) g_{33}^2 + g_{43}^2 \Big].
\end{eqnarray}
\end{widetext}

\section{RG corrections to test verticies}
\label{app:test}

Here, we show the renormalization of test vertices in one-loop order and the processes of obtaining irreducible representations of order parameters.

\subsection{Density-wave with momentum $Q$}

The test vertex term in the Hamiltonian is Eq.~(\ref{eq:HamilDW}), and the vertex correction is Eq. (\ref{eq:test_flow}) with Feynman diagrams in Fig.~\ref{fig:RGtest}. The explicit form of flow equations are
\begin{widetext}
\begin{eqnarray}
    \frac{d\Phi_{\gamma, n \sigma, n \sigma'}}{dy}
    &=& d\bigg[ g_{24} \Phi_{\gamma, n \sigma, n \sigma'} + g_{34} \Phi_{\gamma, n\sigma', n\sigma}^*
    - \sum_{\sigma''} \bigg( g_{14} \Phi_{\gamma, n \sigma'', n \sigma''} + g_{34} \Phi_{\gamma, n \sigma'', n \sigma''}^* \bigg) \delta_{\sigma \sigma'} \nn
    && \ \ \ + g_{21} \Phi_{\gamma, \bar n \sigma, \bar n \sigma'} + g_{31} \Phi_{\gamma, \bar n \sigma', \bar n \sigma}^*
    - \sum_{\sigma''} \bigg( g_{12} \Phi_{\gamma, \bar n \sigma'', \bar n \sigma''} + g_{32} \Phi_{\gamma, \bar n \sigma'', \bar n \sigma''}^* \bigg) \delta_{\sigma \sigma'} \bigg],
\end{eqnarray}
for intra-band CDW order and
\begin{eqnarray}
    \frac{d\Phi_{\gamma, n \sigma, \bar n \sigma'}}{dy} 
    &=& d_1 \bigg[ g_{22} \Phi_{\gamma, n \sigma, \bar n \sigma'} + g_{33} \Phi_{\gamma, \bar n \sigma', n \sigma}^* 
    - \sum_{\sigma''} \bigg( g_{11} \Phi_{\gamma, n \sigma'', \bar n \sigma''} + g_{33} \tilde\Phi_{\gamma, n \sigma'', \bar n \sigma''}^* \bigg) \delta_{\sigma \sigma'} \nn
    && \ \ \ + g_{23} \Phi_{\gamma, \bar n \sigma, n \sigma'} + g_{32} \Phi_{\gamma, \bar n \sigma', n \sigma}^*
    - \sum_{\sigma''} \bigg( g_{13} \Phi_{\gamma, \bar n \sigma'', n \sigma''} + g_{31} \Phi_{\gamma, \bar n \sigma'', n \sigma''}^* \bigg) \delta_{\sigma \sigma'} \bigg].
\end{eqnarray}
\end{widetext}
for inter-band CDW order. Note that the patch indices are not mixed. One can diagonalize spin space with eigenvectors,

\begin{eqnarray}
    \Phi_{\gamma, nn'} &=& \frac{1}{\sqrt 2} ( \Phi_{\gamma, n\uparrow, n'\uparrow} + \Phi_{\gamma, n\downarrow, n'\downarrow}), \nn
    \Phi_{\gamma, nn'}^+ &=& \Phi_{\gamma, n\uparrow, n'\downarrow}, \nn
    \Phi_{\gamma, nn'}^- &=& \Phi_{\gamma, n\downarrow, n'\uparrow}, \nn
    \Phi_{\gamma, nn'}^z &=& \frac{1}{\sqrt 2} ( \Phi_{\gamma, n\uparrow, n'\uparrow} - \Phi_{\gamma, n\downarrow, n'\downarrow}).
\end{eqnarray}
The last three spaces degenerate. For the orbital spaces, there are symmetric or antisymmetric combinations,
\begin{eqnarray}
    \Phi_{s/a} &=& \frac{1}{\sqrt 2} (\Phi_{nn'} \pm \Phi_{n'n})
\end{eqnarray}
where the only orbital indices are written for brevity. Taking real and imaginary parts of density-waves, we have
\begin{eqnarray}
    N^{\rm intra}_{\gamma, s/a} = \Re \frac{1}{\sqrt 2} (\Phi_{\gamma, 11} \pm \Phi_{\gamma, 22}), \nn
    \phi^{\rm intra}_{\gamma, s/a} = \Im \frac{1}{\sqrt 2} (\Phi_{\gamma, 11} \pm \Phi_{\gamma, 22}), \nn
    (S^{\rm intra}_{\gamma, s/a})^A = \Re \frac{1}{\sqrt 2} (\Phi_{\gamma, 11}^A \pm \Psi_{\gamma, 22}^A), \nn
    (\psi^{\rm intra}_{\gamma, s/a})^A = \Re \frac{1}{\sqrt 2} (\Phi_{\gamma, 11}^A \pm \Psi_{\gamma, 22}^A),
\end{eqnarray}
and 
\begin{eqnarray}
    N^{\rm inter}_{\gamma, s/a} = \Re \frac{1}{\sqrt 2} (\Phi_{\gamma, 12} \pm \Phi_{\gamma, 21}), \nn
    \phi^{\rm inter}_{\gamma, s/a} = \Im \frac{1}{\sqrt 2} (\Phi_{\gamma, 12} \pm \Phi_{\gamma, 21}), \nn
    (S^{\rm inter}_{\gamma, s/a})^A = \Re \frac{1}{\sqrt 2} (\Phi_{\gamma, 12}^A \pm \Psi_{\gamma, 21}^A), \nn
    (\psi^{\rm inter}_{\gamma, s/a})^A = \Re \frac{1}{\sqrt 2} (\Phi_{\gamma, 12}^A \pm \Psi_{\gamma, 21}^A),
\end{eqnarray}
with $A = +, -, z$. The corresponding mean-field expectation values in terms of electron fields and coupling constants are listed in Table~\ref{tab:channels}.

\subsection{Superconductivity}

The flow equation for pairing in one-loop is
\begin{widetext}
\begin{eqnarray}
    \frac{d\Delta_{\alpha, n\sigma, n \bar\sigma}}{dy} &=& g_{44} \Delta_{\alpha, n\sigma, n \bar\sigma} + \sum_{\beta \neq \alpha} g_{34} \Delta_{\beta, n\sigma, n\bar\sigma} + g_{43} \Delta_{\alpha, \bar n \sigma, \bar n, \bar\sigma} + \sum_{\beta \neq \alpha} g_{33} \Delta_{\beta, \bar n, \sigma, \bar n, \bar\sigma}, \nn
    \frac{d\Delta_{\alpha, n\sigma, \bar n \sigma'}}{dy} &=& g_{42} \Delta_{\alpha, n\sigma, \bar n, \sigma'} + \sum_{\beta \neq \alpha} g_{32} \Delta_{\beta, n\sigma, \bar n, \sigma'} + g_{41} \Delta_{\alpha, \bar n \sigma, n \sigma'} + \sum_{\beta \neq \alpha} g_{31} \Delta_{\beta, \bar n \sigma, n\sigma'}.
\end{eqnarray}
\end{widetext}
Diagonalizing the patch space, we have $s$- and $d$-like spatial dependence with
\begin{eqnarray}
    \Delta_{s;n\sigma, n\bar\sigma} &=& \frac{1}{\sqrt3} (\Delta_{1,n\sigma, n\bar\sigma} + \Delta_{2,n\sigma, n\bar\sigma} + \Delta_{3,n\sigma, n\bar\sigma}), \nn
    \Delta_{d1;n\sigma, n\bar\sigma} &=& \frac{1}{\sqrt2} (\Delta_{2,n\sigma, n\bar\sigma} - \Delta_{3,n\sigma, n\bar\sigma}), \nn
    \Delta_{d2;n\sigma, n\bar\sigma} &=& \frac{1}{\sqrt6} (2\Delta_{1,n\sigma, n\bar\sigma} - \Delta_{2,n\sigma, n\bar\sigma} - \Delta_{3,n\sigma, n\bar\sigma}). \nn
\end{eqnarray}
For the intra-orbital pairings, the spin should be singlet and the orbital indices can be symmetric or antisymmetric:
\begin{eqnarray}
    \Delta_{B,s}^{\rm intra} &=& \frac{1}{2} (\Delta_{B;1 \uparrow,1 \downarrow} + \Delta_{B;2 \uparrow, 2\downarrow} - \Delta_{B;1 \downarrow,1 \uparrow} - \Delta_{B;2 \downarrow, 2\uparrow}), \nn
    \Delta_{B,a}^{\rm intra} &=& \frac{1}{2} (\Delta_{B;1 \uparrow,1 \downarrow} - \Delta_{B;2 \uparrow, 2\downarrow} - \Delta_{B;1 \downarrow,1 \uparrow} + \Delta_{B;2 \downarrow, 2\uparrow}), \nn
\end{eqnarray}
with $B=s, d1, d2$. For the inter-orbital pairings, the spin can be singlet or triplet corresponding to the orbital triplet or singlet to have total antisymmetric in exchange for electron pairs. So,
\begin{eqnarray}
    \Delta^{\rm inter}_{B,s} &=& \frac{1}{2} (\Delta_{B,1\uparrow, 2\downarrow} + \Delta_{B,2\uparrow, 1\downarrow} - \Delta_{B,1\downarrow, 2\uparrow} - \Delta_{B,2\downarrow, 1\uparrow}), \nn
    (\Delta^{\rm inter}_{B,t})^+ &=& \frac{1}{\sqrt 2}(\Delta_{B,1\uparrow, 2\downarrow} - \Delta_{B,2\uparrow, 1\downarrow}), \nn
    (\Delta^{\rm inter}_{B,t})^- &=& \frac{1}{\sqrt 2}(\Delta_{B,1\downarrow, 2\uparrow} - \Delta_{B,2\downarrow, 1\uparrow}), \nn
    (\Delta^{\rm inter}_{B,t})^z &=& \frac{1}{2}(\Delta_{B,1\uparrow, 2\uparrow} - \Delta_{B,1\downarrow, 2\downarrow} - \Delta_{B,2\uparrow, 1\uparrow} + \Delta_{B,2\downarrow, 1\downarrow}), \nn
\end{eqnarray}
with $B=s, d1, d2$. The corresponding mean-field expectation values in terms of electron fields and coupling constants are listed in Table~\ref{tab:channels}. Note that SC$^{\rm intra}_{d1/d2,a}$ and SC$^{\rm inter}_{d1/d2,t}$ are degenerate and can be formed time-reversal-symmetry-broken orders.

\section{Landau-Ginzberg functional for degenerate SC states}

\subsection{Degenerate spin-singlet SC phase}

The BdG action for degenerate $d$-wave-like orbital-symmetric spin-singlet pair, ($dOS$, $\bar dOS$), is given by
\begin{eqnarray}
    S &=& \int_0^\beta\!d\tau \ \sum_k c_k^\dagger(\partial_\tau + \epsilon_k) c_k 
    + \frac{g}{2} \big( \hat\Delta_1^* \hat\Delta_1 + \hat\Delta_2^* \hat\Delta_2 \big), \nn
\end{eqnarray}
where
\begin{eqnarray}
    \hat\Delta_1 = \frac{1}{2} \sum_k c_{-k} (\lambda^3 \otimes \tau^{0} \otimes i\sigma^2) c_k, \nn
    \hat\Delta_2 = \frac{1}{2} \sum_k c_{-k} (\lambda^8 \otimes \tau^{0} \otimes i\sigma^2) c_k.
\end{eqnarray}
Performing the Hubbard-Stratonovich transformation for $\hat\Delta_1$ and $\hat\Delta_2$, and neglecting the fluctuations of the fields (mean-field approximation), we have
\begin{eqnarray}
    S_{\rm MF} = \frac{1}{2}\int_0^\beta\!d\tau \ \sum_k \psi_k^\dagger \big[-\mathcal G^{-1}_k(\Delta_1, \Delta_2)\big] \psi_k \nn
    - \frac{2}{g} \big( |\Delta_1|^2 + |\Delta_2|^2 \big).
\end{eqnarray}
Here, $\psi_k = (c_k, i\sigma^2 c_{-k}^\dagger)^T$ is Balian-Werthammer spinor,
\begin{eqnarray}
    &&-\mathcal G^{-1}_k (\Delta_1, \Delta_2) = \begin{pmatrix}
        \partial_\tau + \epsilon_k & \sum_{a=1,2} \Delta_a \Lambda_a \\ \sum_{a=1,2} \Delta_a^* \Lambda_a^* & \partial_\tau - \epsilon_{-k}
    \end{pmatrix}, \nn
    &&\hat\Lambda_1 = \frac{1}{2} \lambda^3 \otimes \tau^{0} \otimes i\sigma^2, \ \ \
    \hat\Lambda_2 = \frac{1}{2} \lambda^8 \otimes \tau^{0} \otimes i\sigma^2,
\end{eqnarray}
and $\Delta_a = \big< \hat\Delta_a \big>$. Integrating out $\psi_k$, one can find the mean-field free-energy 
\begin{eqnarray}
    F = -\Tr \log (1-\mathcal G_0 \Delta) + \frac{2}{g} \sum_{a=1,2} |\Delta_a|^2
\end{eqnarray}
where $\Delta = -\mathcal G^{-1} + \mathcal G_0^{-1}$ with $\mathcal G_0^{-1} = -\mathcal G^{-1}(\Delta_a=0)$. Expanding in $\Delta$,
\begin{eqnarray}
    F \approx \frac{2}{g} \sum_{a=1,2} |\Delta_a|^2 + \frac{1}{2} \Tr (\mathcal G_0 \Delta)^2 + \frac{1}{4} \Tr (\mathcal G_0 \Delta)^4.
\end{eqnarray}
Note that the odd term vanishes due to the off-diagonal structure of $\Delta$ in particle-hole space. The quadratic term reads
\begin{eqnarray}
    F^{(2)} &=& \frac{1}{2} \Tr (\mathcal G_0 \Delta)^2 \nn
    &=& \Tr (G_+ G_-) \Tr \sum_{a,b} \Delta_a^* \Delta_b \Lambda_a^* \Lambda_b \nn
    &=& 2\Tr (G_+ G_-) \big( |\Delta_1|^2 + |\Delta_2|^2 \big).
\end{eqnarray}
Here, $G_+ = (-\partial_\tau - \epsilon_k)^-1$ and $G_- = (-\partial_\tau + \epsilon_{-k})^{-1}$. For the quartic order term,
\begin{eqnarray}
    F^{(4)} &=& \frac{1}{4} \Tr (\mathcal G_0 \Delta)^4 \nn
    &=& \frac{1}{2} \Tr (G_+^2 G_-^2) \nn
    &&\times \Tr \sum_{a,b,c,d} \Delta_a^* \Delta_b \Delta_c^* \Delta_d \Lambda_a^* \Lambda_b \Lambda_c^* \Lambda_d \nn
    &=& \frac{1}{4} \Tr (G_+^2 G_-^2) \big[ (|\Delta_1|^2 + |\Delta_2|^2)^2 \nn
    &&+ \frac{1}{3} (\Delta_1 \Delta_2^* - \Delta_1^* \Delta_2)^2 \big].
\end{eqnarray}
Collecting the terms, the free-energy in quartic order in $\Delta$ is
\begin{eqnarray}
    F &=& \bigg[\frac{2}{g} + 2\Tr (G_+ G_-)\bigg] \big(|\Delta_1|^2 + |\Delta_2|^2\big) \nn
    &&+ \frac{1}{4} \Tr (G_+^2 G_-^2) \bigg[ \big(|\Delta_1|^2 + |\Delta_2|^2 \big)^2 \nn
    &&- \frac{4}{3}|\Delta_1|^2 |\Delta_2|^2 \sin^2\varphi \bigg],
\end{eqnarray}
where $\varphi = \arg (\Delta_1/\Delta_2)$. This is the form of free-energy, Eq.~(\ref{eq:LG1}) in the main text. Note that the other degenerate spin-singlet pairs, ($dZS$, $\bar dZS$) and ($dXS$, $\bar dXS$)

Note that the free energy for the degenerate order parameters, $dZS$ SC, 
\begin{eqnarray}
    \Delta_3 &=& \frac{1}{2} \sum_k c_{-k} (\lambda^3 \otimes \tau^3 \otimes i\sigma^2) c_k, \nn
    \Delta_5 &=& \frac{1}{2} \sum_k c_{-k} (\lambda^8 \otimes \tau^3 \otimes i\sigma^2) c_k,
\end{eqnarray}
has the same structure as the $dOS$ SC case.

\subsection{Degenerate spin-triplet SC phase}

The BdG action for degenerate $d$-wave-like inter-orbital-singlet spin-triplet pair, ($diYT$, $\bar diYT$) is given by
\begin{eqnarray}
    S = \int_0^\beta\!d\tau \ \sum_k c_k^\dagger (\partial_\tau + \epsilon_k) c_k + \frac{g}{2}\big( \hat{\bm\Delta}_1^* \cdot \hat{\bm\Delta}_1 + \hat{\bm\Delta}_2^* \cdot \hat{\bm\Delta}_2 \big), \nn
\end{eqnarray}
where
\begin{eqnarray}
    \hat{\bm\Delta}_1 &=& \frac{1}{2} \sum_k c_{-k} (\lambda^3 \otimes \tau^1 \otimes \bm \sigma i\sigma^2) c_k, \nn
    \hat{\bm\Delta}_2 &=& \frac{1}{2} \sum_k c_{-k} (\lambda^8 \otimes \tau^1 \otimes \bm \sigma i\sigma^2) c_k.
\end{eqnarray}
Similar to the spin-singlet case, performing Hubbard-Stratonovich transformation, introducing BW spinor, and integrating out spinor fields yield the following free-energy,
\begin{eqnarray}
    F = -\Tr \log (1-\mathcal G_0 \Delta) + \frac{2}{g} \sum_{a=1,2} |\bm\Delta_a|^2.
\end{eqnarray}
Here, 
\begin{eqnarray}
    \Delta = \begin{pmatrix}
        0 & \sum_a \bm\Delta_a \cdot \bm\Lambda_a \\
        \sum_a \bm\Delta_a \cdot \bm\Lambda_a^* & 0
    \end{pmatrix},
\end{eqnarray}
$\bm\Delta_a = \big< \hat{\bm\Delta}_a \big>$ with $a=1,2$, and
\begin{eqnarray}
    \bm\Lambda_1 = \frac{1}{2} \lambda^3 \otimes \tau^1 \otimes \bm\sigma i\sigma^2, \nn
    \bm\Lambda_2 = \frac{1}{2} \lambda^8 \otimes \tau^1 \otimes \bm\sigma i\sigma^2.
\end{eqnarray}
Expanding in $\Delta$,
\begin{eqnarray}
    F \approx \frac{2}{g} \sum_a |\bm\Delta_a|^2 + \frac{1}{2} \Tr(\mathcal G_0 \Delta)^2 + \frac{1}{4} \Tr(\mathcal G_0 \Delta)^4.
\end{eqnarray}
The quadratic term is
\begin{eqnarray}
    F^{(2)} &=& \frac{1}{2} \Tr(\mathcal G_0 \Delta)^2 \nn
    &=& \Tr(G_+ G_-) \Tr \sum_{a,b} \sum_{i,j} \Delta_a^{i*} \Delta_b^j \Lambda_a^{i*} \Lambda_b^k \nn
    &=& 2\Tr(G_+ G_-) \big(|\bm\Delta_1|^2 + |\bm\Delta_2|^2\big).
\end{eqnarray}
The quartic term is
\begin{eqnarray}
    F^{(4)} &=& \frac{1}{4} \Tr (\mathcal G_0 \Delta)^4 \nn
    &=& \frac{1}{2} \Tr(G_+^2 G_-^2) \nn
    &&\times \Tr \sum_{a,b,c,d} \sum_{i,j,k,l} \Delta_a^{i*} \Delta_b^j \Delta_c^{k*} \Delta_d^l \Lambda_a^{i*} \Lambda_b^j \Lambda_c^{k*} \Lambda_d^l. \nn
\end{eqnarray}
To have non-vanishing traces, we have two cases: (i) $a=b=c=d=1,2$, (ii) two are 1, and the other two are 2 among $a$, $b$, $c$, and $d$. For the case (i), we have
\begin{eqnarray}
    F^{(4,i)} &=& \frac{1}{4} \Tr (G_+^2 G_-^2) \sum_a \big(|\bm\Delta_a|^4 + |\bm\Delta_a^* \times \bm\Delta_a|^2 \big). \nn
\end{eqnarray}
For the case (ii), one can find
\begin{eqnarray}
    F^{(4,ii)} &=& \frac{1}{2} \Tr(G_+^2 G_-^2) \bigg[ \frac{2}{3} |\bm\Delta_1|^2 |\bm\Delta_2|^2 \nn
    && -\frac{1}{6} \big(\bm\Delta_1^2 \bm\Delta_2^{*2} + \bm\Delta_1^{*2} \bm\Delta_2^2\big) \nn
    &&+ \frac{1}{3} \big\{ (\bm\Delta_1 \cdot \bm\Delta_2^*)^2 + (\bm\Delta_1^* \cdot \bm\Delta_2)^2 \big\} \nn
    && + \frac{2}{3} |\bm\Delta_1 \cdot \bm\Delta_2^*|^2 - \frac{2}{3} |\bm\Delta_1 \cdot \bm\Delta_2|^2 \bigg].
\end{eqnarray}
In total, the free energy up to quadratic order in $\Delta$ is
\begin{eqnarray}
    F &=& \bigg[\frac{2}{g} + 2\Tr(G_+ G_-) \bigg] \big(|\bm\Delta_1|^2 + |\bm\Delta|^2 \big) \nn
    && + \frac{1}{4} \Tr(G_+^2 G_-^2) \bigg[ \big( |\bm\Delta_1|^2 + |\bm\Delta_2|^2 \big)^2 \nn
    && + |\bm\Delta_1^* \times \bm\Delta_1|^2 + |\bm\Delta_2^* \times \bm\Delta_2|^2 \nn
    && - \frac{2}{3} |\bm\Delta_1|^2 |\bm\Delta_2|^2 -\frac{1}{3} \big(\bm\Delta_1^2 \bm\Delta_2^{*2} + \bm\Delta_1^{*2} \bm\Delta_2^2\big) \nn
    &&+ \frac{2}{3} \big\{ (\bm\Delta_1 \cdot \bm\Delta_2^*)^2 + (\bm\Delta_1^* \cdot \bm\Delta_2)^2 \big\} \nn
    && + \frac{4}{3} |\bm\Delta_1 \cdot \bm\Delta_2^*|^2 - \frac{4}{3} |\bm\Delta_1 \cdot \bm\Delta_2|^2 \bigg].
\end{eqnarray}
Using the identities,
\begin{eqnarray}
    |\bm\Delta_1 \times \bm\Delta_2|^2 &=& |\bm\Delta_1|^2 |\bm\Delta_2|^2 - |\bm\Delta_1 \cdot \bm\Delta_2^*|^2, \nn
    |\bm\Delta_1 \times \bm\Delta_2^*|^2 &=& |\bm\Delta_1|^2 |\bm\Delta_2|^2 - |\bm\Delta_1 \cdot \bm\Delta_2|^2, \nn
    (\bm\Delta_1 \times \bm\Delta_2^*)^2 &=& \bm\Delta_1^2 \bm\Delta_2^{*2} - (\bm\Delta_1 \cdot \bm\Delta_2^*)^2,
\end{eqnarray}
the free energy can be reformulated as
\begin{eqnarray}
    F &=& \bigg[\frac{2}{g} + 2\Tr(G_+ G_-) \bigg] \big(|\bm\Delta_1|^2 + |\bm\Delta|^2 \big) \nn
    && + \frac{1}{4} \Tr(G_+^2 G_-^2) \bigg[ \big( |\bm\Delta_1|^2 + |\bm\Delta_2|^2 \big)^2 \nn
    && + |\bm\Delta_1^* \times \bm\Delta_1|^2 + |\bm\Delta_2^* \times \bm\Delta_2|^2 \nn
    && + \frac{1}{3} \bigg\{ - 2|\bm\Delta_1|^2 |\bm\Delta_2|^2 + \bm\Delta_1^2 \bm\Delta_2^{*2} + \bm\Delta_1^{*2} \bm\Delta_2^2 \nn
    && - 2(\bm\Delta_1 \times \bm\Delta_2^*)^2 - 2(\bm\Delta_1^* \times \bm\Delta_2)^2 \nn
    && - 4|\bm\Delta_1 \times \bm\Delta_2^*|^2 + 4|\bm\Delta_1 \times \bm\Delta_2|^2 \bigg\} \bigg],
\end{eqnarray}
which is Eq.~(\ref{eq:LG2}) in the main text.

\bibliography{kagome_RG_refs.bib}

\end{document}